\documentclass[comment,  prd, twocolumn, nofootinbib, preprintnumbers, superscriptaddress]{revtex4-2}
\usepackage{amsmath,amssymb,bm,slashed,braket}
\usepackage{graphicx}
\usepackage{epstopdf}
\usepackage{float}
\usepackage{xcolor}
\usepackage{multirow}
\usepackage{dsfont}
\usepackage[colorlinks=true,
            linkcolor=blue,
            urlcolor=blue,
            citecolor=purple,          
            bookmarks=true,
            bookmarksnumbered=true,
            breaklinks=true,
            pdfpagemode=Fullscreen,
            pdfstartview=FitBH]{hyperref}
\usepackage[normalem]{ulem}
\allowdisplaybreaks[4]
\usepackage[capitalise]{cleveref}
\usepackage{orcidlink}
\usepackage{microtype}

\usepackage{physics}
\usepackage[title]{appendix} % 定义附录标记
\usepackage{subcaption}
\usepackage{tikz}
\begin{document}
\title{Breit interaction in elastic scattering of vortex electron on Hydrogen atomic target}
\author{Pengcheng Zhao\,\orcidlink{0000-0001-9211-6016}}
\email{20240264@m.scnu.edu.cn}
\affiliation{State Key Laboratory of Nuclear Physics and
	Technology, Institute of Quantum Matter, South China Normal
	University, Guangzhou 510006, China}
\affiliation{Guangdong Basic Research Center of Excellence for
	Structure and Fundamental Interactions of Matter, Guangdong
	Provincial Key Laboratory of Nuclear Science, Guangzhou
	510006, China}

\begin{abstract}

The interaction of electron vortex beams with atoms is fundamental to their applications, yet existing studies of elastic scattering have considered only the Coulomb interaction.
The role of the Breit interaction, which represents the leading relativistic correction to the Coulomb potential between two electrons, has not been systematically examined.
In this paper, we study the elastic scattering of a vortex electron by a hydrogen atomic target, with emphasis on the role of the Breit interaction.
Comparisons are made between two situations: the pure Coulomb contribution and the pure Breit contribution.
Our results show that the Breit interaction dominates the scattering process when the total angular momentum of the vortex electron is larger than a certain threshold, which challenges the conventional notion that the Breit interaction makes little contribution to electron scattering for low-$Z$ atoms.
This phenomenon arises primarily because a vortex electron with large angular momentum possesses a substantial magnetic moment, and the Breit interaction precisely describes the interaction between magnetic moments.
The threshold can be adjusted by changing the parameters of the initial vortex electron.
Moreover, we further compare the single-atom, mesoscopic, and macroscopic target cases, and show that the target size significantly influences the observability of the Breit-induced effects, with a smaller target size being favorable for observing this phenomenon.
Vortex electrons add a new tool to the well-established electron-atom scattering tool box.
Our results are important for applications of vortex electrons: they show that, when large total angular momentum is involved, the Breit interaction must be taken into account in calculations.

\end{abstract}
\maketitle
\section{Introduction}
The wave function of a free electron can exhibit spatial structures far richer than that of a plane wave.
One example is the vortex state, whose wavefronts possess a helical phase structure, with a phase singularity extending along the propagation direction, around which the probability current flows in a spiral manner.
Such states are characterized by their topological charge (or winding number) and carry an intrinsic orbital angular momentum (OAM), which is fundamentally distinct from spin angular momentum (SAM) \cite{allen1992orbital}.
Since the 2010s, electron vortex beams (EVBs) have been successfully realized in transmission electron microscopes (TEM) with kinetic energies of \(200\) keV or \(300\) keV \cite{uchida2010nature, shiloh2014Ultra, verbeeck2010nature, grilloprx2014, schatprl2012, bechenp2014}.
Their topological charge can be as large as thousands (positive or negative) \cite{mafaapl2017,tavabiapl2022}.
The realization of EVBs has stimulated extensive interest in their interactions with matter. In the field of electron microscopy, EVBs have been proposed for applications such as nanoparticle manipulation \cite{verbeekam2013}, chiral structure characterization \cite{juchprb2015}, high-resolution magnetic imaging \cite{rusznc2016}, among others \cite{ivanovprl2013, juchpra2016, schatultra2017, koneprr2023, bliokhpr2017, ivanovppnp2022}.

Most applications of EVBs rely on their interaction with atoms, which makes it necessary to study EVB scattering by atomic targets.
Several works have focused on this investigation.
For example, Rutherford scattering with an initial Bessel vortex electron and a single atomic target has been studied using a screened Coulomb potential \cite{boxempra2014}.
Later, inelastic scattering in the same scattering configuration was investigated \cite{boxempra2015}.
A relativistic version of Mott scattering has also been discussed \cite{serbopra2015, ivanovpra2023}.
Moreover, scattering of vortex electron wave packets has been studied for different impact parameters \cite{karlovetspra2015, karlovetspra2017}.
Laguerre-Gaussian-type vortex wave packets have also been considered for elastic scattering \cite{sheremetarxiv2024}.
All of the studies listed above are carried out within the framework of the first Born approximation.
There are also works that provide results beyond this framework \cite{koshpra2018}.

However, in the studies above, only the Coulomb interaction has been considered for elastic scattering.
In the relativistic case, only the initial electron is described by a relativistic state.
The relativistic correction originating from the Breit interaction has not been systematically examined.
The Breit interaction describes the leading relativistic corrections to the Coulomb potential between two electrons, including magnetic interactions and retardation effects \cite{breitpr1929, breitpr1932, bethebook2013}.
It is generally recognized that the influence of the Breit interaction diminishes rapidly for small nuclear charge numbers, and for low-\(Z\) atoms, such as hydrogen, it can be safely neglected \cite{grantbook2007}.
However, when the incident electron carries a large OAM, its magnetic moment is significantly enhanced, and the contribution of the Breit interaction may become non-negligible, or even dominant in certain kinematic regions. This possibility has not yet been explored.

In this paper, we study the elastic scattering of a vortex electron carrying OAM by a hydrogen atomic target, with emphasis on the role of the Breit interaction.
Kinetic energy of initial vortex electron is chosen to be hundreds of keV.
We adopt the Bessel vortex state to describe the initial electron, describe the hydrogen ground state by the four-component spinor wave function of the Dirac equation, and include both the Coulomb and Breit terms in the interaction potential.
By calculating the effective differential cross section as a function of the scattering angle \(\theta_p\) and effective cross section as a function of the topological charge \(m_v\) of the initial vortex electron, we reveal the dominant effect of the Breit interaction in specific parameter regions, which breaks with the conventional notion that the Breit interaction has little contribution to the electron scattering process for low-\(Z\) atoms.
This phenomenon arises primarily because a vortex electron with large OAM possesses a substantial magnetic moment, and the Breit interaction precisely describes the interaction between magnetic moments.
We also examine how parameters of vortex electron influence the effect induced by Breit interaction.
In addition, we systematically compare the results for the single-atom target, mesoscopic target, and macroscopic target cases, and discuss how the target size influences the effects originating from the Breit contribution.

The paper is organized as follows:

Section \ref{section two} presents the theoretical calculations of transition amplitudes and effective differential cross section.

Section \ref{section three} provides multiple results and analyzes the observed phenomena.

Section \ref{section four} concludes the study.

Throughout this paper, three-dimensional vectors are denoted in boldface, e.g., $\vb{r}$, whereas two-dimensional vectors are represented by a letter with an arrow, e.g., $\vec{r}$.
$\hbar=c=1$ is used.

\section{Calculations}\label{section two}
\subsection{Potential and transition amplitude}
For elastic scattering of an electron on a hydrogen atom, the interaction potential is
\begin{align}
    V(\mathbf{r},\mathbf{r}')=V_{\mathrm{Coulomb}}(\mathbf{r},\mathbf{r}')+V_{\mathrm{Breit}}(\mathbf{r},\mathbf{r}'),
\end{align}
where
\begin{align}
    V_{\mathrm{Coulomb}}(\mathbf{r},\mathbf{r}')=\frac{e^2}{4\pi \epsilon_0}\left(-\frac{1}{|\mathbf{r}|}+\frac{1}{R}\right)
\end{align}
and
\begin{align}
    V_{\mathrm{Breit}}(\mathbf{r},\mathbf{r}')=-\frac{e^2}{4\pi \epsilon_0}\frac{1}{2R}\left[ \boldsymbol{\alpha}\cdot \boldsymbol{\alpha}'+\frac{(\boldsymbol{\alpha}\cdot \mathbf{R})(\boldsymbol{\alpha}'\cdot \mathbf{R})}{R^2}\right],
\end{align}
where $\mathbf{r}'=(r',\theta',\phi')$ is the position of the atomic electron, $\mathbf{r}=(r,\theta,\phi)$ is the position of the incident electron, $R=|\mathbf{R}|=|\mathbf{r}-\mathbf{r}'|$, and $\boldsymbol{\alpha}$ and $\boldsymbol{\alpha}'$ are the Dirac $\alpha$ matrices acting on the state spaces of the two electrons, respectively.
Here, $e$ is the electron charge and $\epsilon_0$ is the permittivity of vacuum.
To calculate the scattering process, the atomic wave function as a solution of the Dirac equation is necessary.

The transition amplitude is $\mathcal{M}=\mathcal{M}_{\mathrm{C}}+\mathcal{M}_{\mathrm{B}}$ with
\begin{align}
    \mathcal{M}_{\mathrm{C}}&=\int d^3 \mathbf{r}\, d^3\mathbf{r}'\,
    \psi^+_f(\mathbf{r})\varphi^+_f(\mathbf{r}')
    V_{\mathrm{Coulomb}}(\mathbf{r}, \mathbf{r}')
    \varphi_i(\mathbf{r}')\psi_i(\mathbf{r}),\label{mcint}\\
    \mathcal{M}_{\mathrm{B}}&=\int d^3 \mathbf{r}\, d^3\mathbf{r}'\,
    \psi^+_f(\mathbf{r})\varphi^+_f(\mathbf{r}')
    V_{\mathrm{Breit}}(\mathbf{r}, \mathbf{r}')
    \varphi_i(\mathbf{r}')\psi_i(\mathbf{r}).\label{mvint}
\end{align}
$\psi_{i,f}$ are the initial and final states of the scattered electron with momenta $\mathbf{k}$ and $\mathbf{p}$, respectively, and $\varphi_{i,f}$ are the initial and final atomic states.

\subsection{Vortex electron}
Vortex particles are described by vortex solutions of the equations of motion (such as the Schrödinger, Dirac, and Klein--Gordon equations) expressed in cylindrical coordinates. These vortex states exhibit a singularity line along their propagation direction, around which the probability current flows in a spiral manner, resembling the current in an ideal solenoid. They are characterized by their intrinsic angular momentum, which is proportional to the topological charge (or winding number) $m_v$ appearing in the phase factor $\mathrm{e}^{im_v \phi}$.
For scalar case, it is orbital angular momentum (OAM).
While for spinor case, it is total angular momentum (TAM).
The winding number $m_v$ takes discrete values, either integers (scalar case) or half-integers (spinor case).
Each vortex particle carries an intrinsic angular momentum of $\hbar m_v$, a property fundamentally distinct from spin. Vortex electrons can be described either in a scalar framework (non-relativistic case) or using a spinor formalism (relativistic case), as outlined below.
\subsubsection{Scalar case}
A general expression for a vortex state is given by
\begin{align}
    \psi^V(\mathbf{r}) = g(r_\perp, z) \, \mathrm{e}^{im_v\phi_r},
\end{align}
where $\mathbf{r} = (x, y, z)$ is the three-dimensional position vector, $r_\perp = \sqrt{x^2 + y^2}$ is the transverse radial coordinate, and $\phi_r = \arctan(y/x)$ is the corresponding azimuthal angle.

The simplest example of a vortex state is the Bessel vortex state, which will be used throughout this work. A Bessel vortex state is characterized by three quantum numbers: the energy $E$, the topological charge $m_v$, and the transverse quantum number $\kappa$. It is explicitly given by
\begin{align}\label{bessel}
    \psi^B(\mathbf{r}) = \int \frac{d^2 \vec{k}_\perp}{(2\pi)^2} \, a_{\kappa m_v}(\vec{k}_\perp) \, \psi^P(\mathbf{r}),
\end{align}
where
\begin{align}
    a_{\kappa m_v}(\vec{k}_\perp) = (-i)^{m_v} \, \mathrm{e}^{im_v\phi_k} \sqrt{\frac{2\pi}{\kappa}} \, \delta(|\vec{k}_\perp| - \kappa)
\end{align}
is the momentum-space amplitude, $\psi^P$ denotes a plane-wave state with momentum $\mathbf{k}$, and $\vec{k}_\perp$ is the transverse component of $\mathbf{k}$. In the scalar (i.e., non-relativistic) case considered here, $m_v$ takes integer values.

\subsubsection{Spinor case}
For the spinor case, we simply substitute the plane-wave spinor electron state
\begin{align}
    \psi^P(\vb r) = u(\vb k, \vb s_k) \, \mathrm{e}^{i\vb k \cdot \vb r}
\end{align}
into Eq.~\eqref{bessel}. Here, $u(\vb k, \vb s_k)$ is a Dirac spinor with momentum $\vb k$ and polarization vector $\vb s_k$. In this work, we employ helicity eigenstates to describe the Dirac spinor, as they are particularly convenient for calculations.
A Dirac spinor with momentum $\vb k$ and helicity $\lambda_k$ is given by
\begin{align}
    u(\vb k,\lambda_k) = \frac{1}{\sqrt{2E_k}}
    \begin{pmatrix}
        \sqrt{E_k + m_e} \; \omega(\vb n_k, \lambda_k) \\
        2\lambda_k \sqrt{E_k - m_e} \; \omega(\vb n_k, \lambda_k)
    \end{pmatrix},
\end{align}
where $E_k$ is the electron energy, $m_e$ its mass, and $\vb n_k$ the direction of $\vb k$. The spinor $\omega(\vb n, \lambda)$ is an eigenstate of the helicity operator $\hat\Lambda(\vb n) = \frac{1}{2}\hat{{\sigma}}\cdot\vb n$ with eigenvalue $\lambda = \pm \frac{1}{2}$, where $\hat{{\sigma}}$ denotes the Pauli matrices.
For their formulas with different helicity, see APPENDIX.~\ref{appendix a}.
These vortex states are common eigenstates of both the TAM operator and the helicity operator. Their TAM equals $\hbar m_v$, with $m_v$ taking half-integer values in the spinor case.

\subsection{Hydrogen atom wave function based on Dirac equation}
The wave function of the hydrogen atom, as a solution of the Dirac equation, is described by a four-component spinor.
For the ground state $1S_{\frac{1}{2}}$ with $n=1$, $j=\frac{1}{2}$, $l=0$, and $m=\frac{1}{2}$, the hydrogen wave function is \cite{bethebook2013}
\begin{align}\label{ground state wave function}
   \varphi_{1\frac{1}{2}0\frac{1}{2}}(\mathbf{r})= \begin{pmatrix}
    \frac{1}{\sqrt{\pi}}a_0^{-\frac{3}{2}}\sqrt{\frac{1+\epsilon_{1}}{\Gamma(2\gamma_0 +1)}}\mathrm{e}^{-\frac{1}{2}R}R^{\gamma _0-1}\\
    0\\
     \frac{i}{\sqrt{\pi}}a_0^{-\frac{3}{2}}\sqrt{\frac{1-\epsilon_{1}}{\Gamma(2\gamma_0 +1)}}\mathrm{e}^{-\frac{1}{2}R}R^{\gamma _0-1}\cos \theta\\
     \frac{i}{\sqrt{\pi}}a_0^{-\frac{3}{2}}\sqrt{\frac{1-\epsilon_{1}}{\Gamma(2\gamma_0 +1)}}\mathrm{e}^{-\frac{1}{2}R}R^{\gamma _0-1}\sin \theta \,\mathrm{e}^{i\phi_r}
    \end{pmatrix},
\end{align}
with the energy eigenvalue $E_{1\frac{1}{2}}=m_e\sqrt{\frac{\sqrt{1-\alpha ^2}}{\alpha ^2+\sqrt{1-\alpha ^2}}}$.
Here, $a_0$ is the Bohr radius,
$m_e$ is the electron mass,
$\alpha\approx 1/137$ is the fine-structure constant,
$\epsilon_1=E_{1\frac{1}{2}}/m_e$, $\gamma_0 =\sqrt{1-\alpha ^2}$and $R=\frac{2r}{a_0}$.

\subsection{Calculation for initial plane wave electron}
\subsubsection{Transition amplitude of Coulomb interaction for elastic scattering of electron on atoms}
Calculation of transition amplitude for Coulomb interaction begins with Eq.~\eqref{mcint}.
The integral is straightforward, and we directly give the result:
\begin{align}\label{mc}
    \mathcal{M}_{\mathrm{C}}=\frac{e^2}{\epsilon_0 q^2}u^+_f(\mathbf{p},\lambda_p)u_i(\mathbf{k},\lambda_k)\left[ F_0(\mathbf{q})-1\right],
\end{align}
where
\begin{align}\label{f0}
    F_0(\mathbf{q})=\int d^3\mathbf{r}' \,\varphi_f^+(\mathbf{r}')\varphi_i(\mathbf{r}')\mathrm{e}^{-i\mathbf{q}\cdot \mathbf{r}'}.
\end{align}
Here, $\mathbf{k}$ is the momentum of the incident electron, $\mathbf{p}$ is the momentum of the scattered electron, and $\mathbf{q}=\mathbf{p}-\mathbf{k}$ is the momentum transfer.

For the ground state $1S_{\frac{1}{2}}$, inserting Eq.~\eqref{ground state wave function} into Eq.~\eqref{f0} and Eq.~\eqref{mc} and we obtain
\begin{align}
    \mathcal{M}_{\mathrm{C}}=-\frac{e^2}{\epsilon_0}u^+_f(\mathbf{p},\lambda_p)u_i(\mathbf{k},\lambda_k)\left[ \frac{(8+q^2a_0^2)a_0^2}{(4+q^2a_0^2)^2}+O(\alpha ^2)\right].
\end{align}

The Dirac spinor product is $u^+_f(\mathbf{p},\lambda_p)u_i(\mathbf{k},\lambda_k)=N_D\,\omega^+(\mathbf{n}_p,\lambda_p)\omega(\mathbf{n}_k,\lambda_k)$
where $N_D=\delta_{\lambda_k, \lambda_p}+\frac{m_e}{E_k}\delta_{-\lambda_k,\lambda_p}$
and
\begin{align}
    \omega^+(\mathbf{n}_p,\lambda_p)\omega(\mathbf{n}_k,\lambda_k)=\sum_{\rho=\pm \frac{1}{2}}d_{\rho \lambda_p}^{(1/2)}(\theta_p)\,d_{\rho \lambda_k}^{(1/2)}(\theta_k)\,\mathrm{e}^{-i\rho\phi_{kp}},
\end{align}
where $\phi_{kp}=\phi_k-\phi_p$.
Here $d^{(1/2)}_{\rho\lambda}(\theta_n) = \delta_{\rho,\lambda} \cos\frac{\theta_n}{2} - 2\rho \,\delta_{\rho,-\lambda} \sin\frac{\theta_n}{2}$ is the Wigner $d$-function.

\subsubsection{Transition amplitude of Breit interaction for elastic scattering of electron on atoms}

Calculation of transition amplitude for Breit interaction begins with Eq.~\eqref{mvint}.
The Breit potential is
\begin{align}
    V_{\mathrm{Breit}}(\mathbf{R})=\int \frac{d^3\mathbf{w}}{(2\pi)^3}\,\mathrm{e}^{i\mathbf{w} \cdot \mathbf{R}}\,\tilde{V}_{\mathrm{Breit}}(\mathbf{w}),
\end{align}
where
\begin{align}
    \tilde{V}_{\mathrm{Breit}}(\mathbf{w})
=-\frac{e^2}{\epsilon_0w^2}\left[ \boldsymbol{\alpha} \cdot \boldsymbol{\alpha}^{\prime}-\frac{(\boldsymbol{\alpha} \cdot \mathbf{w})(\boldsymbol{\alpha} ^{\prime} \cdot \mathbf{w})}{w^2}\right].
\end{align}
With this Fourier transform, we obtain the transition amplitude after evaluating the integrals:
\begin{align}
    \mathcal{M}_{\mathrm{B}}=-\frac{e^2}{\epsilon_0q^2}(\mathcal{A}-\mathcal{B}),
\end{align}
where
\begin{align}\label{ab}
    \mathcal{A}=(u_f^+\boldsymbol{\alpha}\, u_i)\cdot \langle\varphi_f(\mathbf{r}')\rvert\,\boldsymbol{\alpha} ^{\prime}\,\mathrm{e}^{-i\mathbf{q} \cdot \mathbf{r}^{\prime}}\lvert\varphi_i(\mathbf{r}')\rangle
\end{align}
and
\begin{align}
    \mathcal{B}=\frac{(u_f^+\boldsymbol{\alpha} \cdot \mathbf{q}\, u_i)\,\langle\varphi_f(\mathbf{r}')\rvert\,\boldsymbol{\alpha}^{\prime} \cdot \mathbf{q}\,\mathrm{e}^{-i\mathbf{q} \cdot \mathbf{r}^{\prime}}\lvert\varphi_i(\mathbf{r}')\rangle}{q^2}.
\end{align}
\paragraph{Calculation of $\cal A$}
For an initial plane-wave electron with momentum \(\mathbf{k}=(k,\theta_k,\phi_k)\) and helicity \(\lambda_k\), and a final plane-wave electron with momentum \(\mathbf{p}=(p,\theta_p,\phi_p)\) and helicity \(\lambda_p\), the momentum modulus is the same for elastic scattering, so we obtain
\begin{align}
    \Big[ u_f^+\boldsymbol{\alpha}\, u_i\Big]_j=(\lambda_k+\lambda_p)\frac{p}{E_k}\,\omega^+(\mathbf{p},\lambda_p)\,\sigma _j\,\omega (\mathbf{k},\lambda_k).
\end{align}
It is obvious that \(a_i\) vanishes for helicity-flipping processes.
That is, \(\mathcal{A}\) is nonzero only if \(\lambda_k=\lambda_p=\lambda=\pm \frac{1}{2}\).
For the ground state \(1S_{\frac{1}{2}}\), inserting Eq.~\eqref{ground state wave function} into Eq.~\eqref{ab}, expanding it at \(\alpha=0\),  and we obtain
\begin{align}
    {\cal A}&=\Bigg[ \frac{-16\lambda p a}{ E_k (4+q^2 a^2)^2}\alpha +O(\alpha ^3)\Bigg] \nonumber\\
    &\quad \times \Big[ \cos \frac{\theta_p}{2}\sin \frac{\theta_k}{2}(p_{\perp}{\rm e}^{i\lambda \phi_{kp}}-k_{\perp}{\rm e}^{-i\lambda \phi_{kp}})\nonumber\\
    &\quad -\cos \frac{\theta_k}{2}\sin \frac{\theta_p}{2}(p_{\perp}{\rm e}^{-i\lambda \phi_{kp}}-k_{\perp}{\rm e}^{i\lambda \phi_{kp}})\Big] .
\end{align}
\paragraph{Calculation of $\cal B$}
Since initial and final electron states are both solution of Dirac equation, we directly have $(u_f^+\vb \alpha  u_i)\cdot \vb q=(E_p-E_k)u_f^+u_i=0$
which results with ${\cal B}\equiv 0$.

\subsection{Calculations for initial vortex electron case}
For Coulomb interaction, helicity-flipping process is far smaller than helicity-conserving process.
And for Breit interaction, helicity-flipping process is forbidden.
Thus we only consider helicity-conserving process in the following.
\subsubsection{Coulomb interaction}
The transition amplitude for an initial Bessel vortex electron is simply the azimuthal average of the plane-wave case, weighted by the winding factor $\mathrm{e}^{im_v\phi_k}$ for each plane-wave component.
For \(\lambda_p=\lambda_k=\lambda=\pm \frac{1}{2}\), the transition amplitude for pure Coulomb interaction is
\begin{align}\label{mvc}
    \mathcal{M}_{\mathrm{VC}}&\approx N_C\int \frac{d \phi_k}{2\pi}\,\mathrm{e}^{im_v\phi_{kp}}\frac{8+q^2a_0^2}{(4+q^2a_0^2)^2}\nonumber\\
    &\quad \times \Big[ \cos \frac{\theta_p}{2}\cos \frac{\theta_k}{2}\mathrm{e}^{-i\lambda \phi_{kp}}+\sin \frac{\theta_p}{2}\sin \frac{\theta_k}{2}\mathrm{e}^{i\lambda \phi_{kp}}\Big],
\end{align}
where $N_C=-4\pi \alpha a_0^2(-i)^{m_v}\sqrt{\frac{\kappa}{2\pi}}{\rm e}^{im_v\phi_{p}}$.

\subsubsection{Breit interaction}
As in the pure Coulomb interaction case, the transition amplitude for the pure Breit interaction is
\begin{align}\label{mvb}
    {\cal M}_{\rm VB}&\approx N_B\int \frac{d \phi_{k}}{2\pi}{\rm e}^{im_v\phi_{kp}}\frac{1}{(4+q^2 a^2)^2q^2}\nonumber\\
    &\quad \times \Big[ \cos \frac{\theta_p}{2}\sin \frac{\theta_k}{2}(p_{\perp}{\rm e}^{i\lambda \phi_{kp}}-k_{\perp}{\rm e}^{-i\lambda \phi_{kp}})\nonumber\\
    &\quad -\cos \frac{\theta_k}{2}\sin \frac{\theta_p}{2}(p_{\perp}{\rm e}^{-i\lambda \phi_{kp}}-k_{\perp}{\rm e}^{i\lambda \phi_{kp}})\Big] 
\end{align}
where $N_B=\frac{64\pi \lambda a_0\alpha ^2p}{  E_k}(-i)^{m_v}\sqrt{\frac{\kappa}{2\pi}}{\rm e}^{im_v\phi_{p}}$.

\subsubsection{Macroscopic and Mesoscopic target cases}
In realistic experiments, it is difficult to prepare a single target
atom in its ground state.
A realistic target has a finite size, which may be either macroscopic
or mesoscopic.
In the former case, the target is sufficiently large that it can be
treated as infinite in extent.
In the latter case, the target size is small, typically on the
nanometer scale.
We consider a homogeneous target with an atomic number density of
$1/(\pi R_m^2)$, where $R_m$ is the radius of the target.
The transition amplitude is
\begin{align}\label{mes}
    |\mathcal{M}_{R}|^2&=\int d^2\vec{b}\,
    \frac{d^2\vec{k}_{\perp}}{(2\pi)^2}
    \frac{d^2\vec{k}^{\prime}_{\perp}}{(2\pi)^2}\,
    \mathcal{M}(\mathbf{k})\,\mathcal{M}^*(\mathbf{k}^{\prime})\,\nonumber\\
    &\quad \times a_{\kappa m_v}(\vec{k}_{\perp})\,
    a^*_{\kappa m_v}(\vec{k}^{\prime}_{\perp})\,
    \frac{\mathrm{e}^{i(\vec{k}^{\prime}_{\perp}-\vec{k}_{\perp})
    \cdot \vec{b}}}{\pi R_m^2}.
\end{align}

For a macroscopic target, we evaluate the integral in
Eq.~\eqref{mes} by taking $R_m\to\infty$.
The integral is straightforward and yields
\begin{align}\label{macro}
     |\mathcal{M}_{R}|^2\propto \int d\phi_k\,
     |\mathcal{M}(\mathbf{k})|^2.
\end{align}
This result is independent of $m_v$, implying that the TAM effect is averaged out and vanishes for a macroscopic target.
Consequently, the macroscopic target case is not considered further; in the following discussion, we focus only on the mesoscopic target case.

\subsection{effective differential cross section}
For collisions involving initial Bessel vortex states, the conventional
definition of a cross section is not directly applicable, as discussed
in Ref.~\cite{karlovetspra2017}.
Instead, we use the effective differential cross section:
\begin{align}\label{efdcs}
    \frac{d\hat {\sigma}}{d\Omega}=\frac{  E_k^2}{2\pi  \kappa \cos \theta_k}|{\cal M}|^2.
\end{align}
In the condition that $m_v=\pm \frac{1}{2}$ and $\kappa \rightarrow 0$, it tends to differential cross section for initial plane wave case with spin $\pm \frac{1}{2}$.
The derivation of this formula is presented in APPENDIX.~\ref{appendix a}.

\section{Results and Discussions}\label{section three}
We consider a central collision between a vortex electron and an
atomic target.
Both the Coulomb and Breit potential operators commute with the TAM operator $\hat{J}_z$ defined about the collision
axis.
Both the initial vortex electron state and the atomic state are
eigenstates of $\hat{J}_z$.
For elastic scattering, in which the atomic target remains in its
ground state, the scattered electron must also be in an eigenstate of
$\hat{J}_z$ with the same TAM as the initial vortex electron.
Consequently, the final-state distribution is always rotationally
symmetric about the propagation axis of the initial vortex electron.
These statements are consistent with the transition amplitudes given
in Eqs.~\eqref{mvc} and~\eqref{mvb}, in which the $\phi_p$
dependence enters solely through the global phase factor
$\mathrm{e}^{im_v\phi_p}$.
Thus, the integral on azimuthal angle $\phi_p$ for effective differential cross section in Eq.~\eqref{efdcs} just give a constant $2\pi$.
After the integral, we get the effective differential cross section
\begin{align}
    \frac{d\hat {\sigma}}{d\theta_p}=\frac{  E_k^2}{ \kappa \cos \theta_k}|{\cal M}|^2.
\end{align}
and the effective cross section
\begin{align}
    \hat{\sigma}=\int ^{\pi}_0d\theta_p \frac{  E_k^2}{ \kappa \cos \theta_k}|{\cal M}|^2.
\end{align}

In this section, we present the effective differential cross section as a function of the scattering angle $\theta_p$ of the scattered electron and the effective cross section as a function of the TAM $m_v$ of the initial vortex electron.
Three cases will be presented.
The first is the case in which only the Coulomb interaction contributes to the transition amplitude $\cal M$, which we denote the ``Coulomb only'' case.
The second is the case in which only the Breit interaction contributes to the transition amplitude $\cal M$, which we denote the ``Breit only'' case.
The third is the case in which both the Coulomb and Breit interactions contribute, which we denote the ``Coulomb+Breit'' case.
For convenience, we normalize the effective cross section to the result for the reference case $m_v=\frac{1}{2}$ of the ``Coulomb+Breit'' case and define the relative effective cross section as
\begin{equation}
    W_R(m_v)=\frac{\hat{\sigma}(m_v)}
    {\hat{\sigma}(1/2)}.
\end{equation}

Since the macroscopic target result is independent of the initial TAM,
we focus in the following on the single-atom and mesoscopic target
cases.

\subsection{$\theta_p$ dependence of effective differential cross section in single atomic target case}
\begin{figure}[h!]
    \centering
        \centering
        \includegraphics[width=0.45\linewidth]{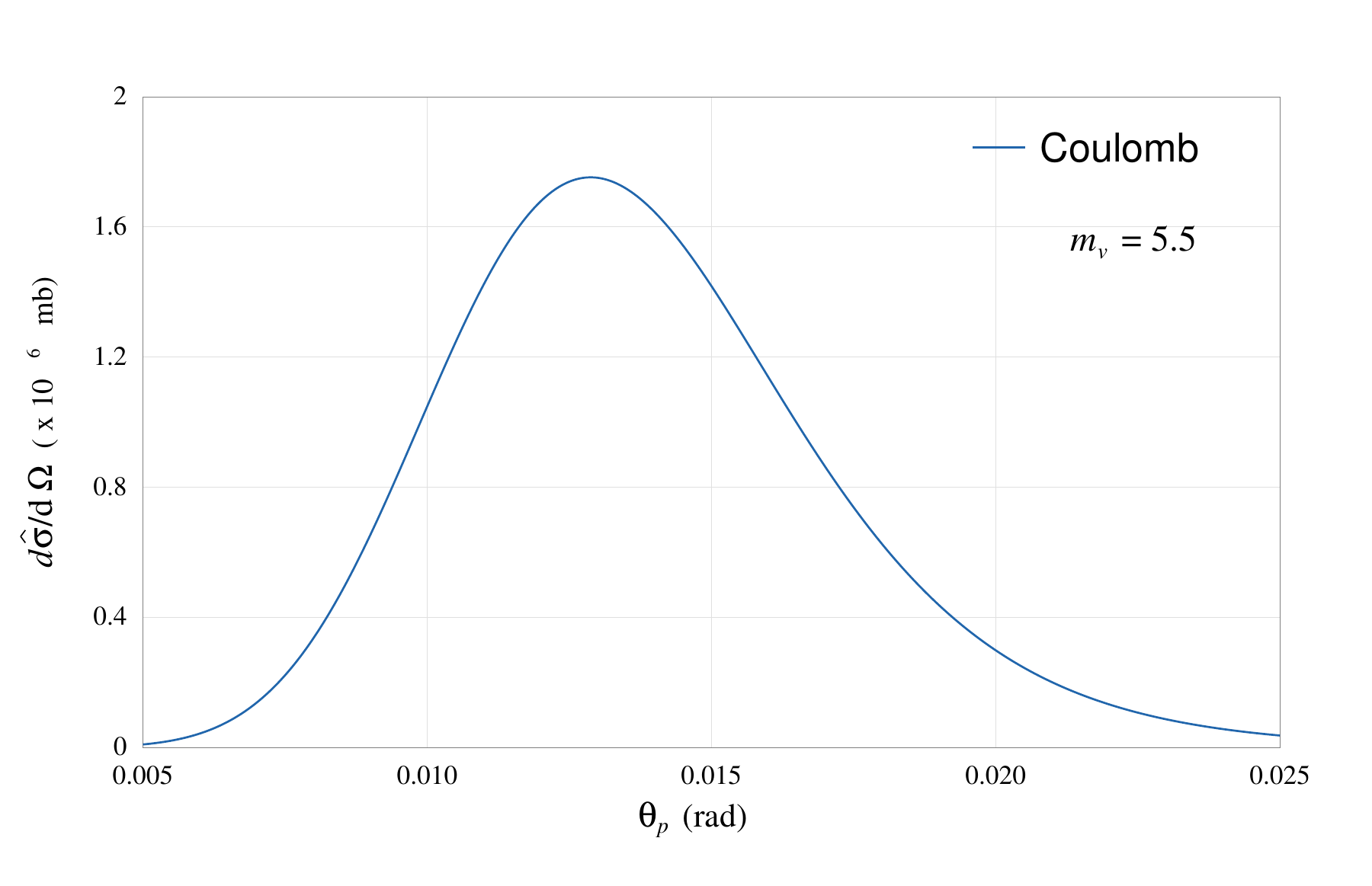}
        \includegraphics[width=0.45\linewidth]{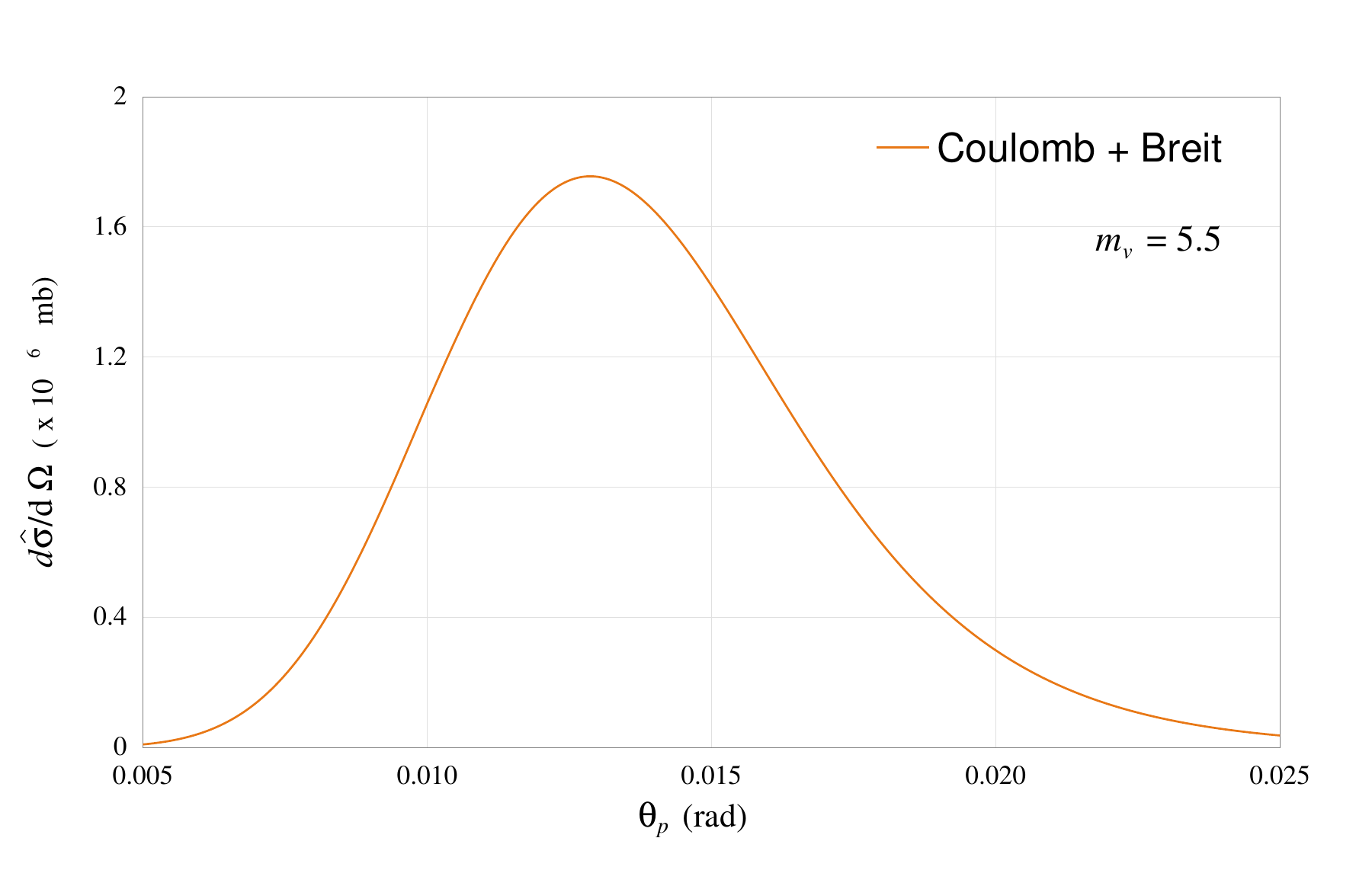}
        \includegraphics[width=0.45\linewidth]{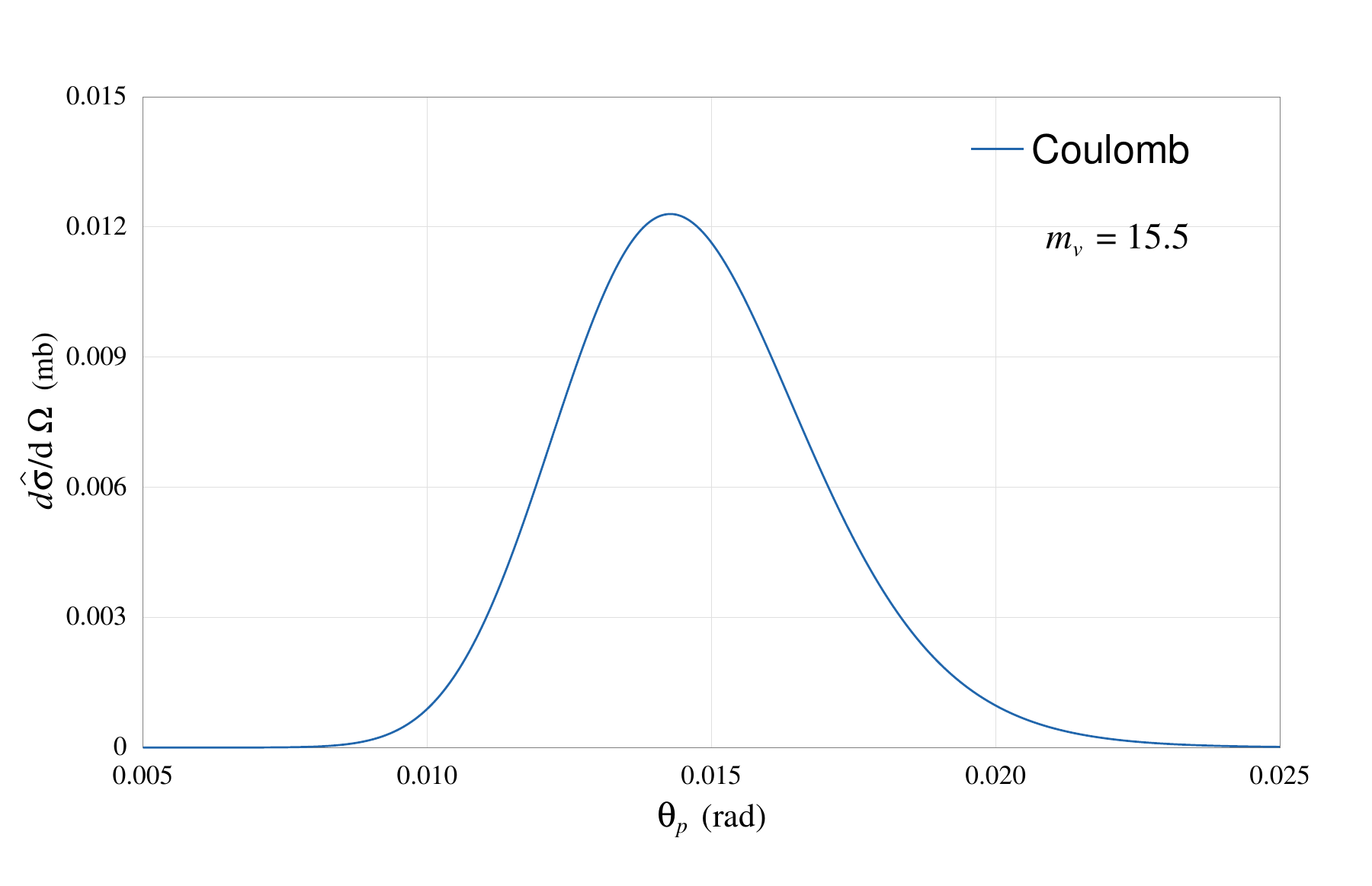}
        \includegraphics[width=0.45\linewidth]{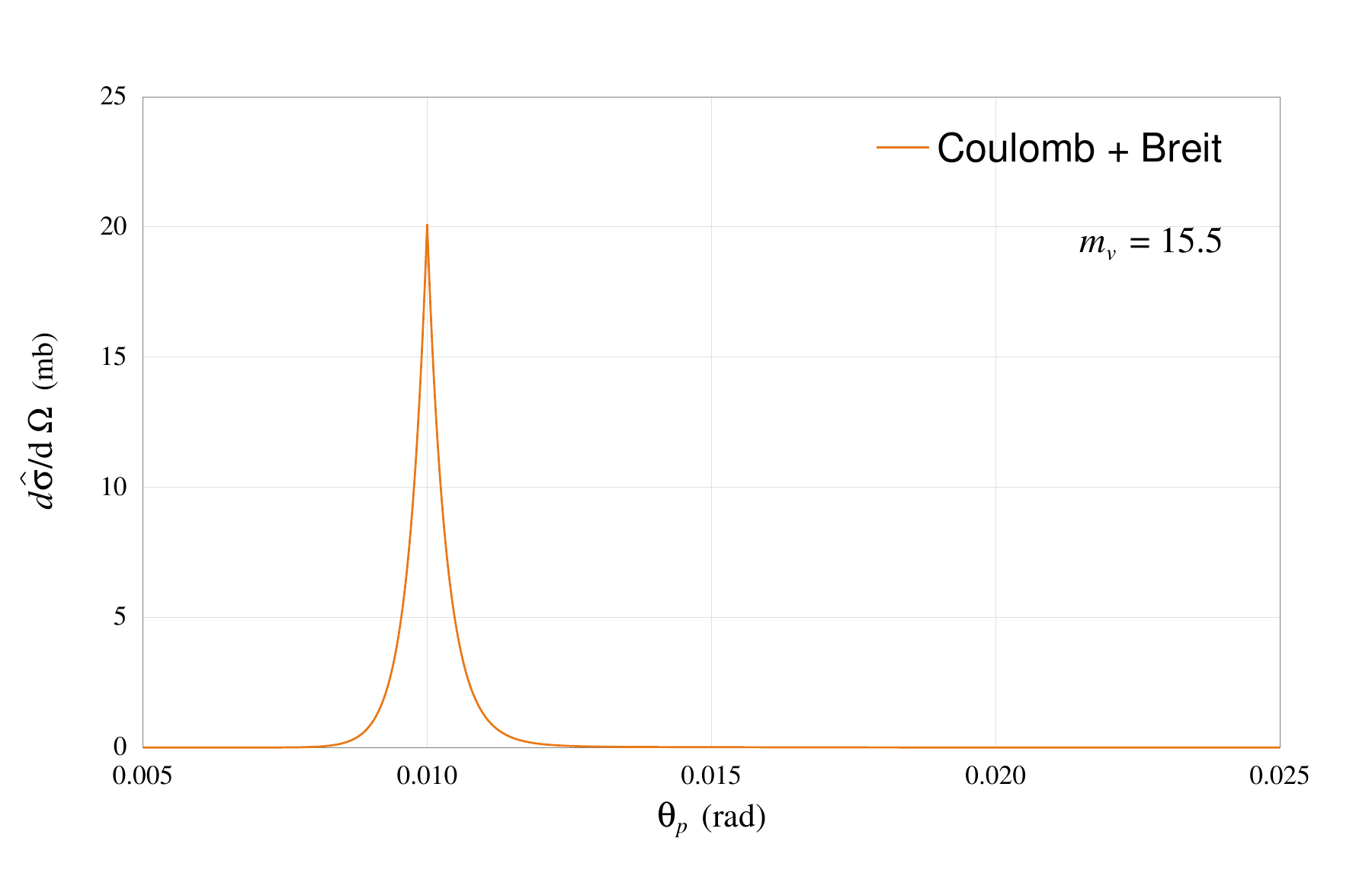}

    \caption{\raggedright Relative effective cross section \(W_R(m_v)\). Dependence on \(\theta_p\) in the single-atom target case. The parameters for the initial vortex electron are  \(E_{\mathrm{kin}}=300~\mathrm{keV}\) and \(\theta_k=0.01\) rad. The helicities are \(\lambda_k=\lambda_p=\frac{1}{2}\). Left panels: pure Coulomb contribution; Right panels: sum of the Coulomb and Breit contributions; Upper panels: $m_v=5.5$; Lower panels: $m_v=15.5$. }
    \label{fig:thetapdep}
\end{figure}
We examine how the effective differential cross section depends on the
scattering angle $\theta_p$ for two different initial TAM values:
$m_v=5.5$ and $m_v=15.5$.
Two cases are compared: the ``Coulomb only'' case, in which only the
Coulomb potential contributes, and the ``Coulomb+Breit'' case, in
which both the Coulomb and Breit potentials contribute.
The results are presented in FIG.~\ref{fig:thetapdep}.
The kinetic energy and conical angle of the initial vortex electron are
$E_{\mathrm{kin}}=300$~keV and $\theta_k=0.01$~rad, respectively.
The helicities of both the incident and scattered electrons are set to
$\frac{1}{2}$.

For the ``Coulomb only'' case, the distributions for both $m_v=5.5$
and $m_v=15.5$ peak around an angle slightly larger than the
conical angle $\theta_k=0.01$.
For $m_v=5.5$, it is $\theta_p\approx 0.013$ rad.
While for $m_v=15.5$, it is $\theta_p\approx 0.014$ rad.
This behavior can be understood as follows.
The initial electron is described by a Bessel vortex state, whose
plane-wave components all carry a fixed polar angle $\theta_k$.
Since the differential cross section for plane-wave electron scattering
peaks in the forward direction, the peak scattering angle for Bessel
vortex electron scattering must lie near $\theta_p=\theta_k$.
The higher the energy of the initial electron, the sharper the peak.
The slight shift of the peak to $\theta_p > \theta_k$ arises from
destructive interference near the propagation axis of the initial
vortex electron.
This destructive interference is most pronounced in the exact forward
direction, where the transition amplitude vanishes.

When both Coulomb and Breit contributions are included, the situation
changes markedly.
For $m_v=5.5$, the distribution coincides with the ``Coulomb
only'' result at all scattering angle $\theta_p$.
This indicates that the Coulomb contribution dominates for all
scattering angles.
For $m_v=15.5$, the distribution differs dramatically from the
``Coulomb only'' result: it peaks sharply at $\theta_p=\theta_k$ with
a magnitude exceeding the Coulomb-only value by three orders of
magnitude.
This demonstrates that the Breit contribution dominates the scattering
process at this larger TAM.

The TAM $m_v$ plays a crucial role in determining the dominant contribution in the scattering process.
To further elucidate how $m_v$ governs the relative importance of the Coulomb and Breit contributions, we present the $m_v$ dependence of the effective cross section in the following subsection.

\subsection{$m_v$ dependence of effective cross section in single atomic target case}

\begin{figure}[h!]
    \centering
        \centering
        \includegraphics[width=0.45\linewidth]{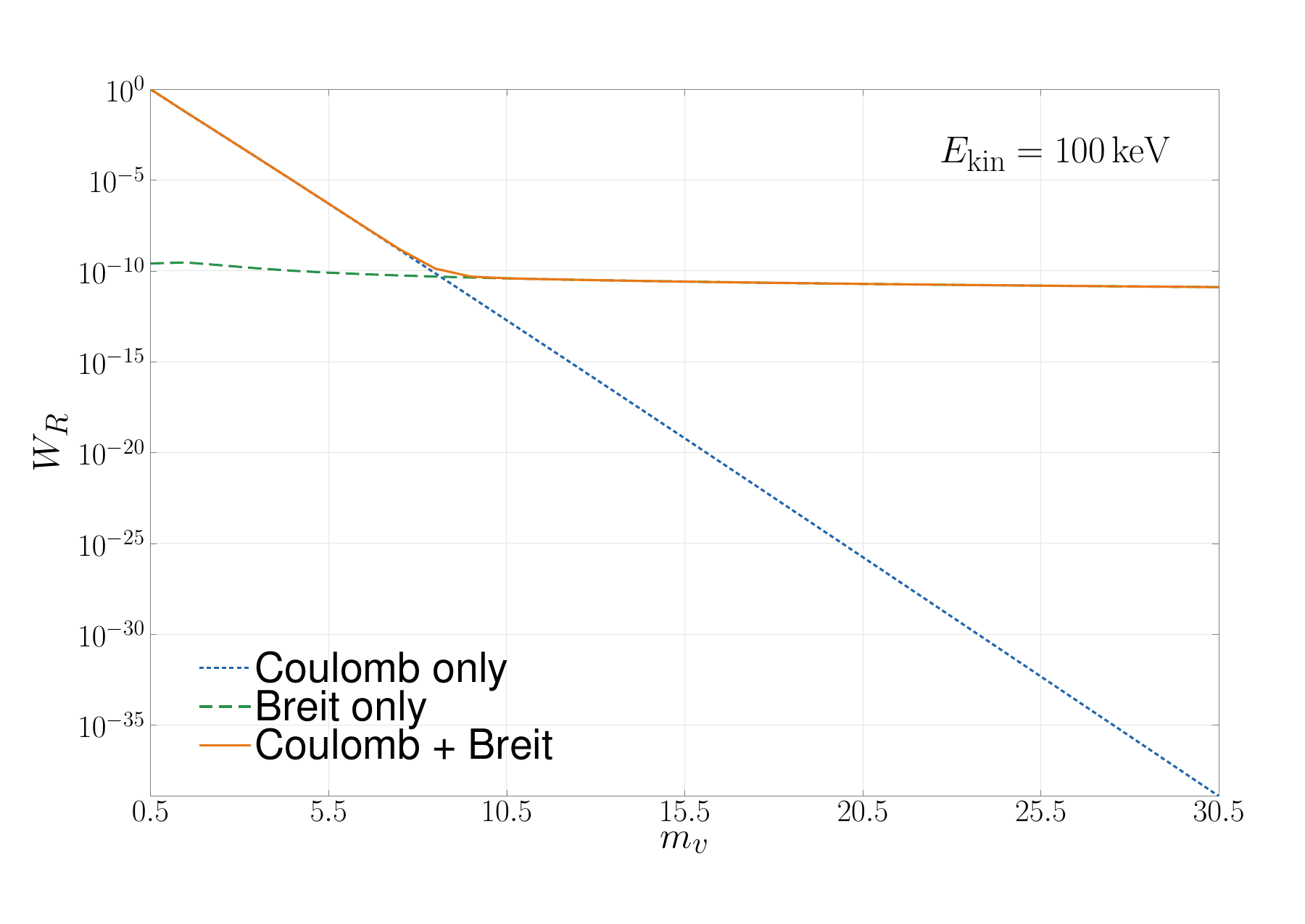}
        \includegraphics[width=0.45\linewidth]{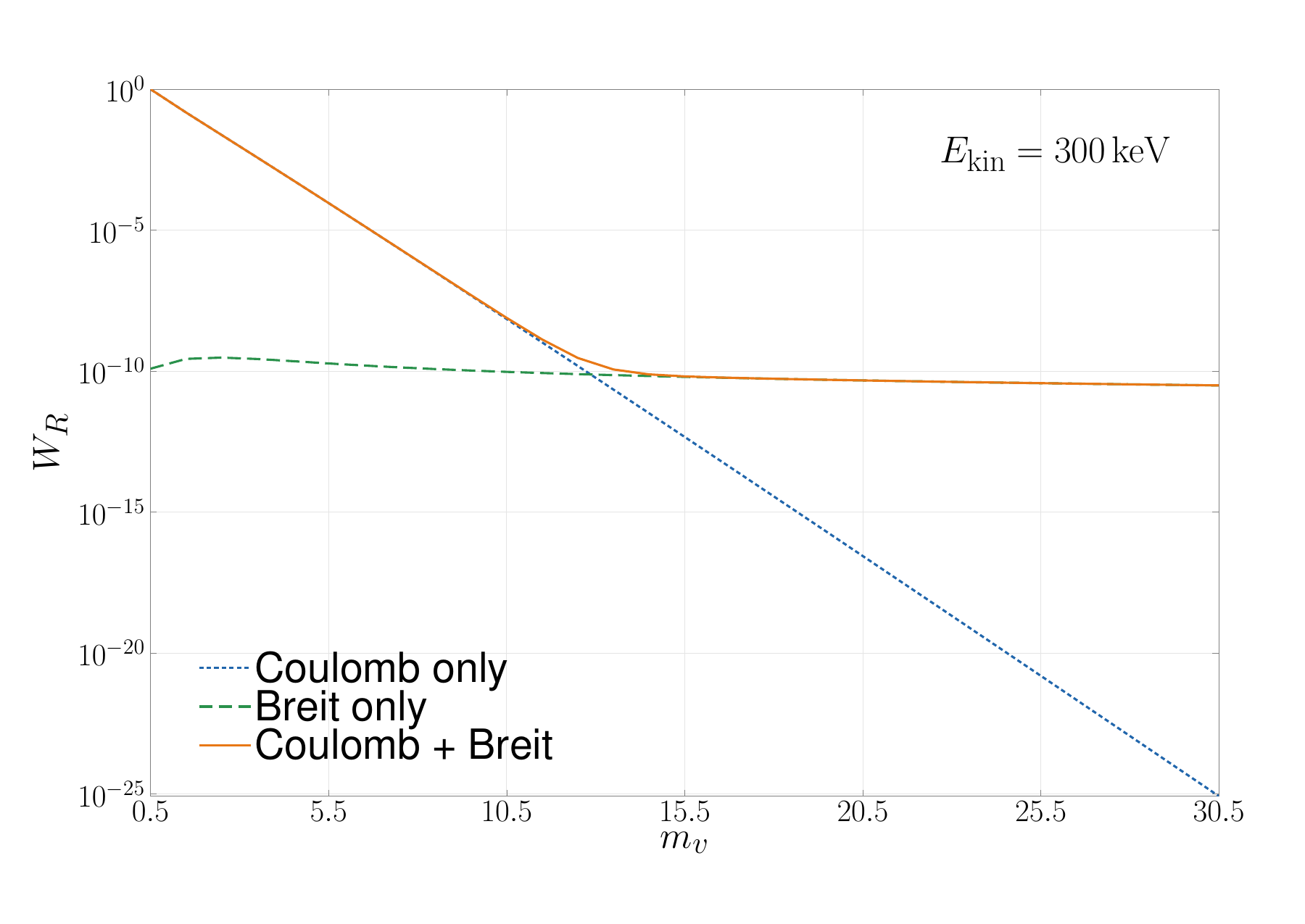}
        \includegraphics[width=0.45\linewidth]{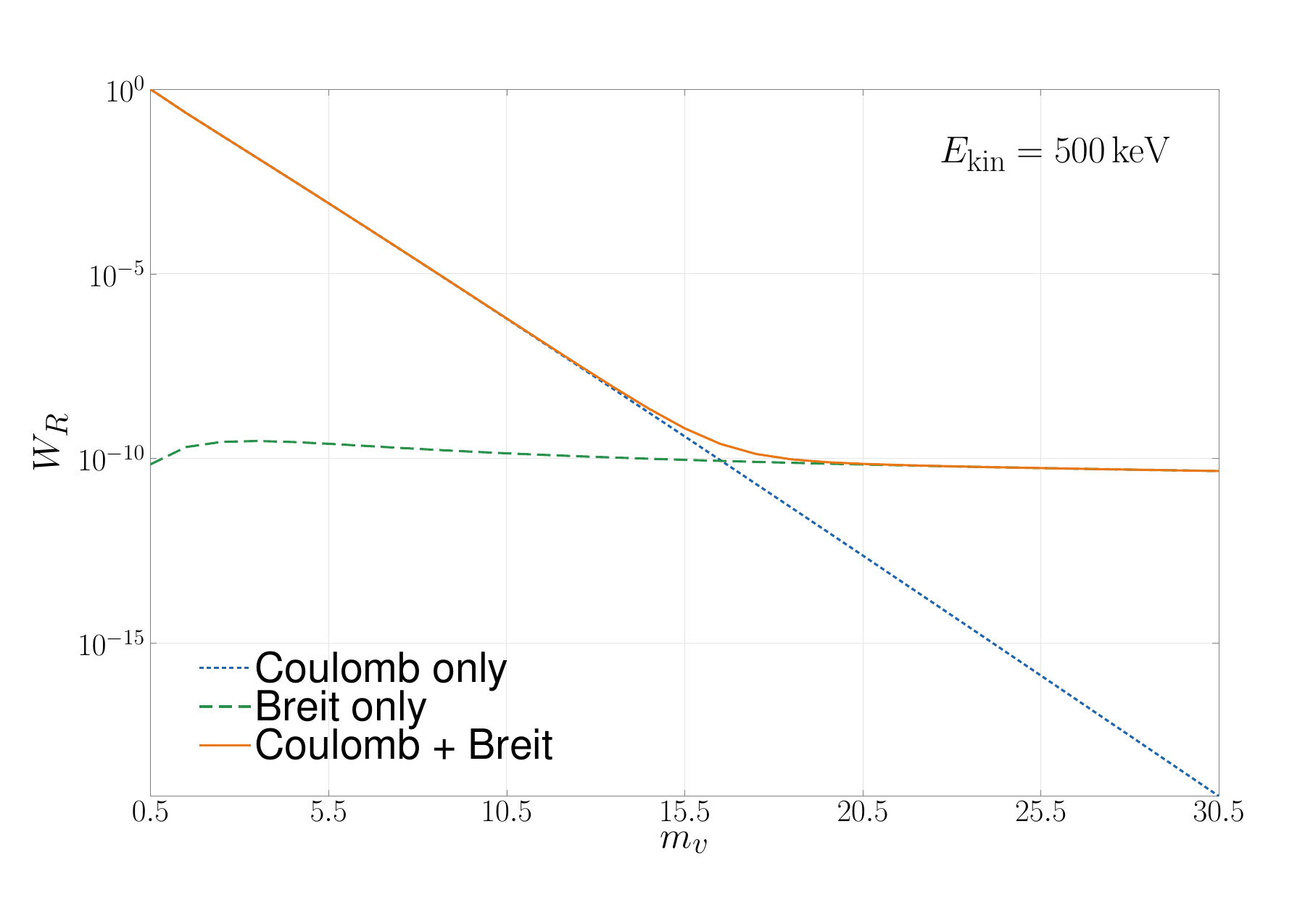}
        \includegraphics[width=0.45\linewidth]{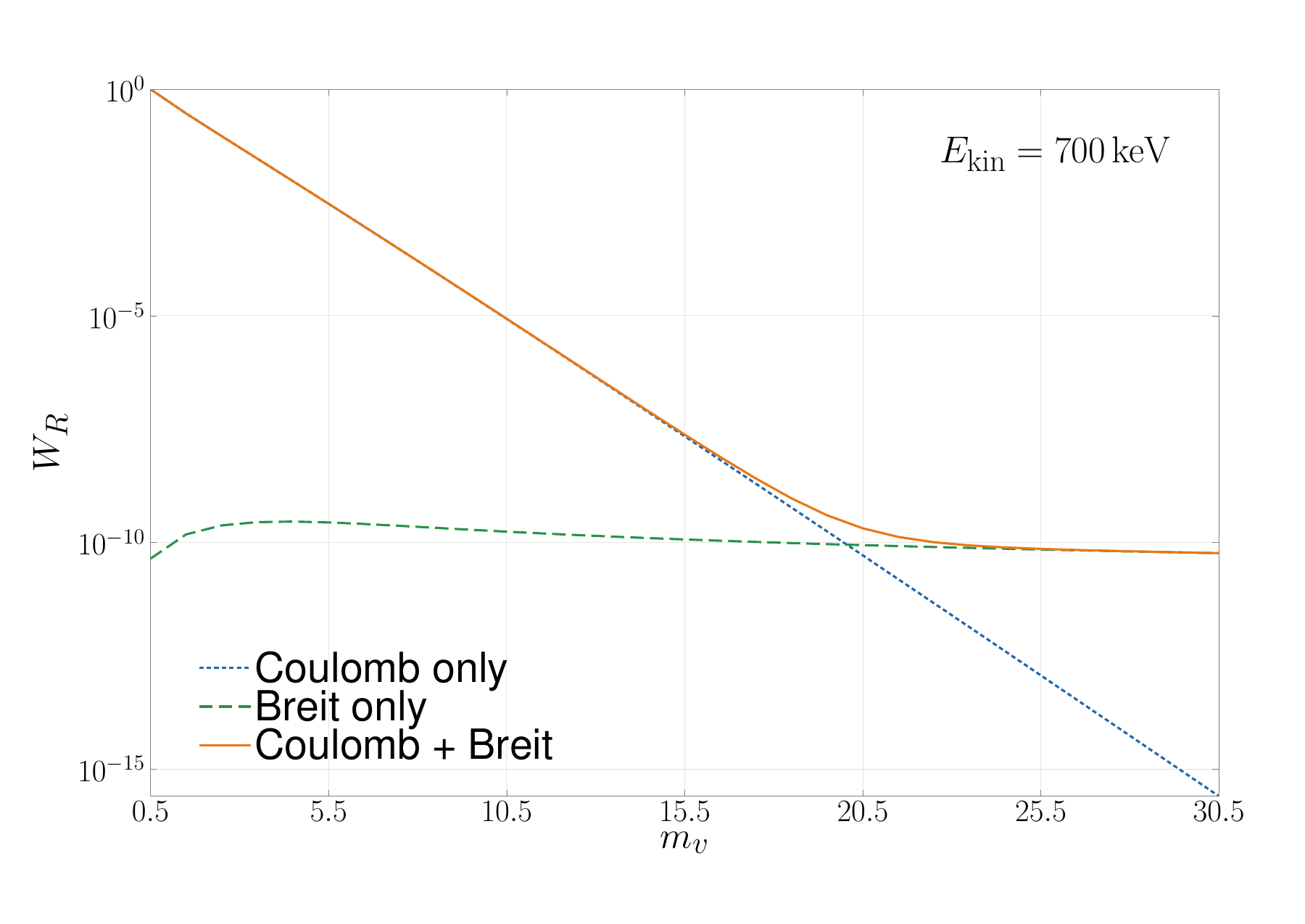}

    \caption{\raggedright Relative effective cross section $W_R(m_v)$ and its dependence on $m_v$ for the single-atom target case.
The kinetic energies of the initial vortex electron are $E_{\mathrm{kin}}=100~\mathrm{keV}$, $300~\mathrm{keV}$, $500~\mathrm{keV}$, and $700~\mathrm{keV}$ respectively.
The conical angle of the initial vortex electron is $\theta_k=0.01$ rad.
The helicities are $\lambda_k=\lambda_p=\frac{1}{2}$.}
    \label{fig:mvdep_Ekin}
\end{figure}

\begin{figure}[h!]
    \centering
        \centering
        \includegraphics[width=0.45\linewidth]{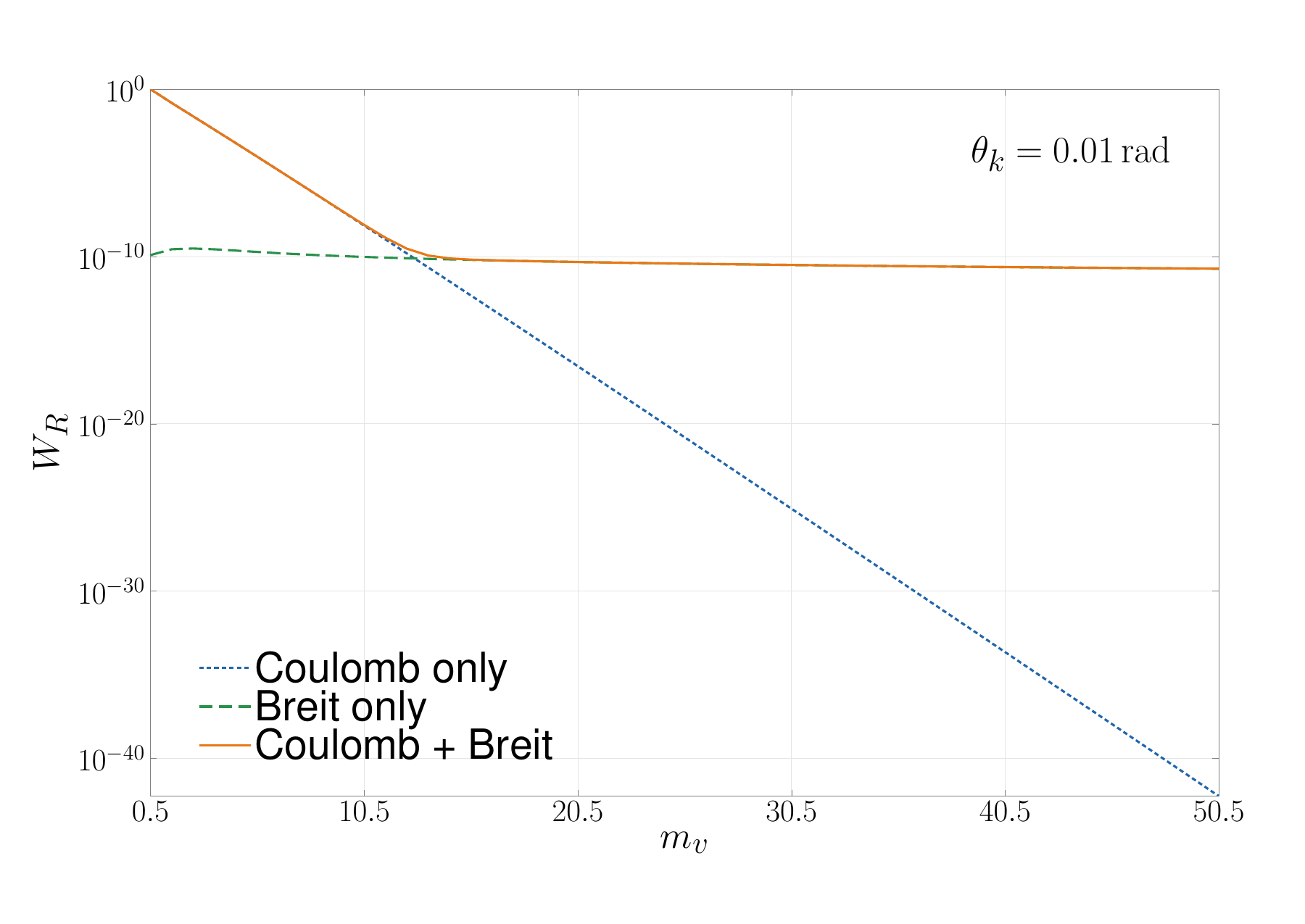}
        \includegraphics[width=0.45\linewidth]{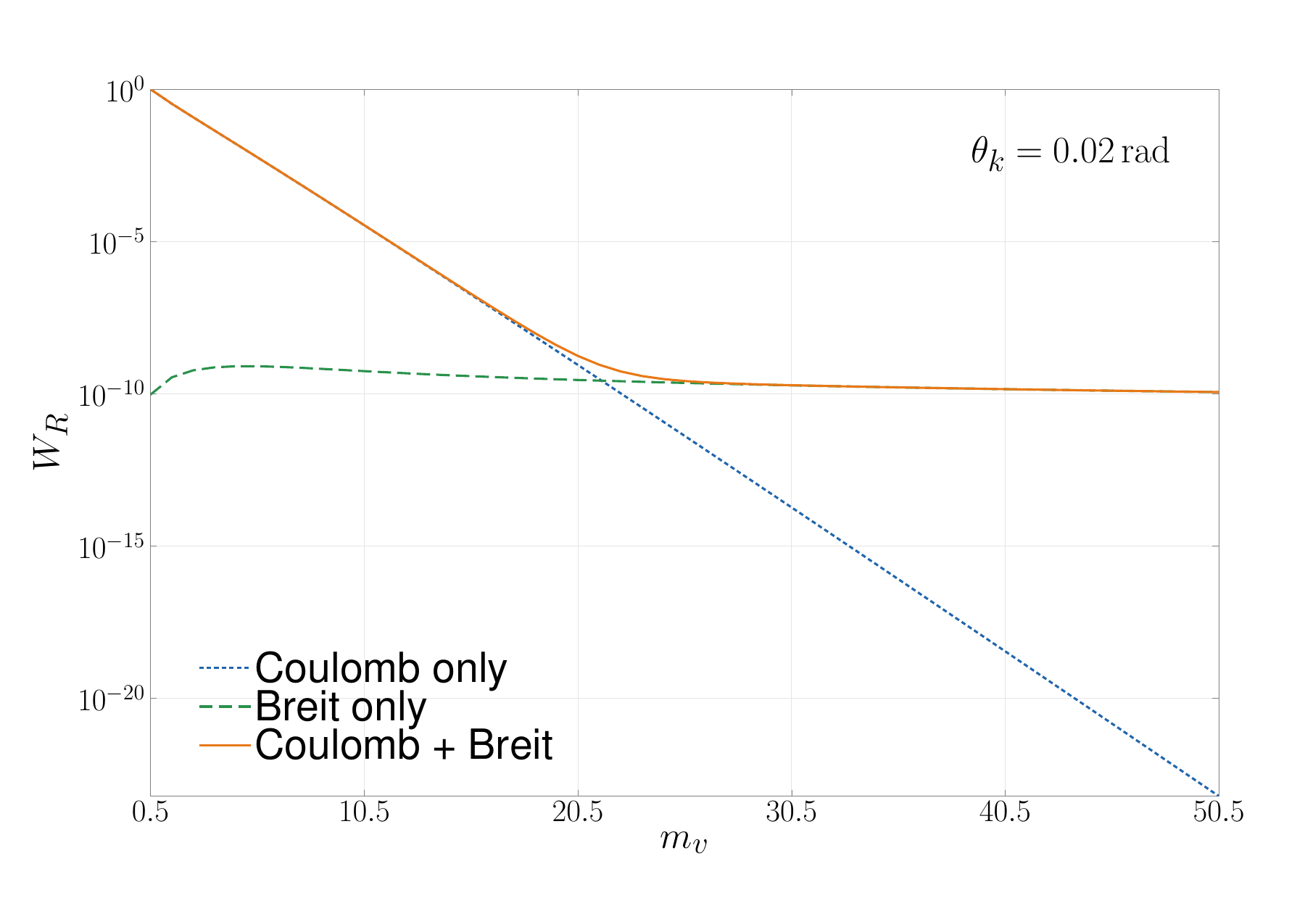}
        \includegraphics[width=0.45\linewidth]{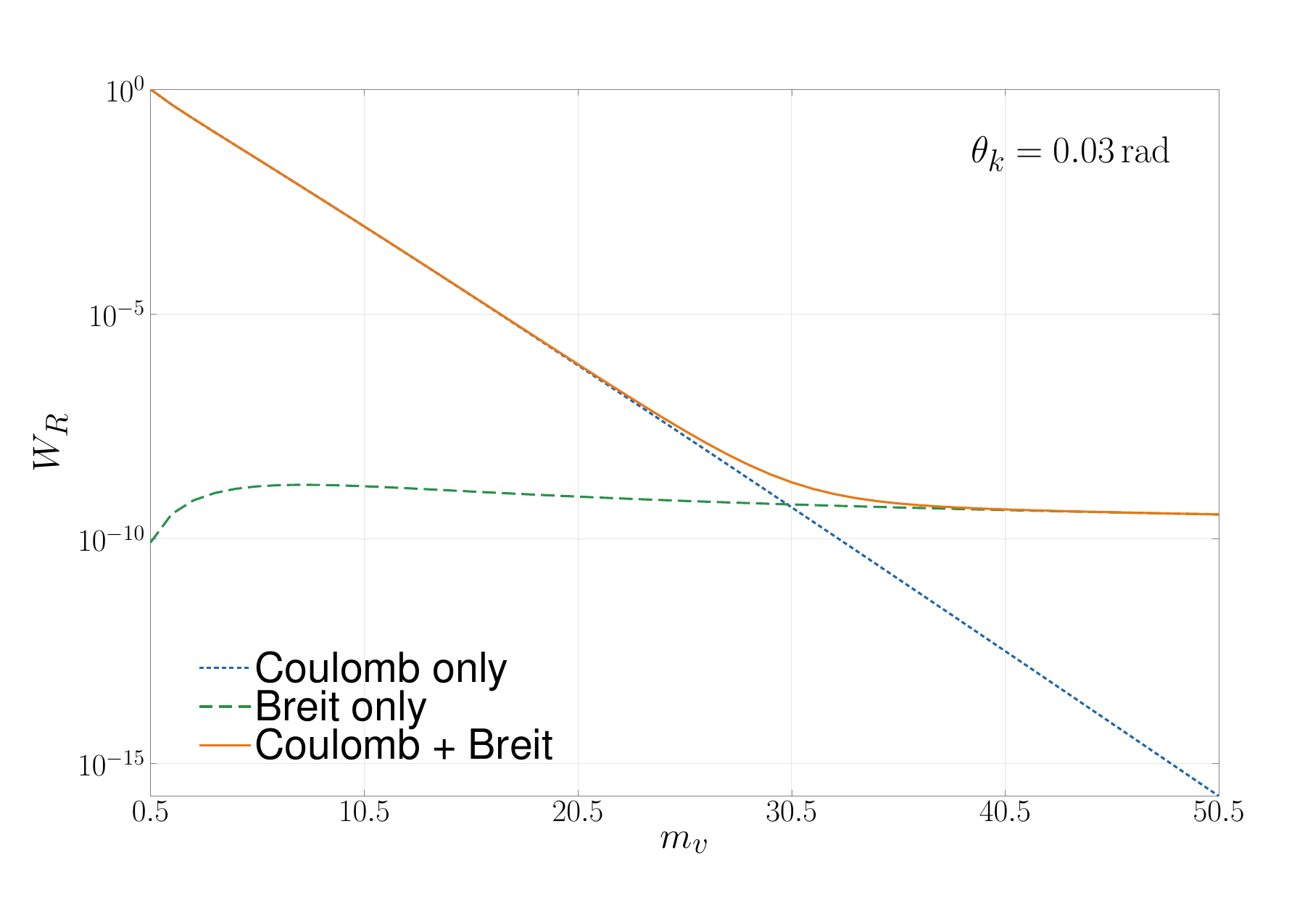}
        \includegraphics[width=0.45\linewidth]{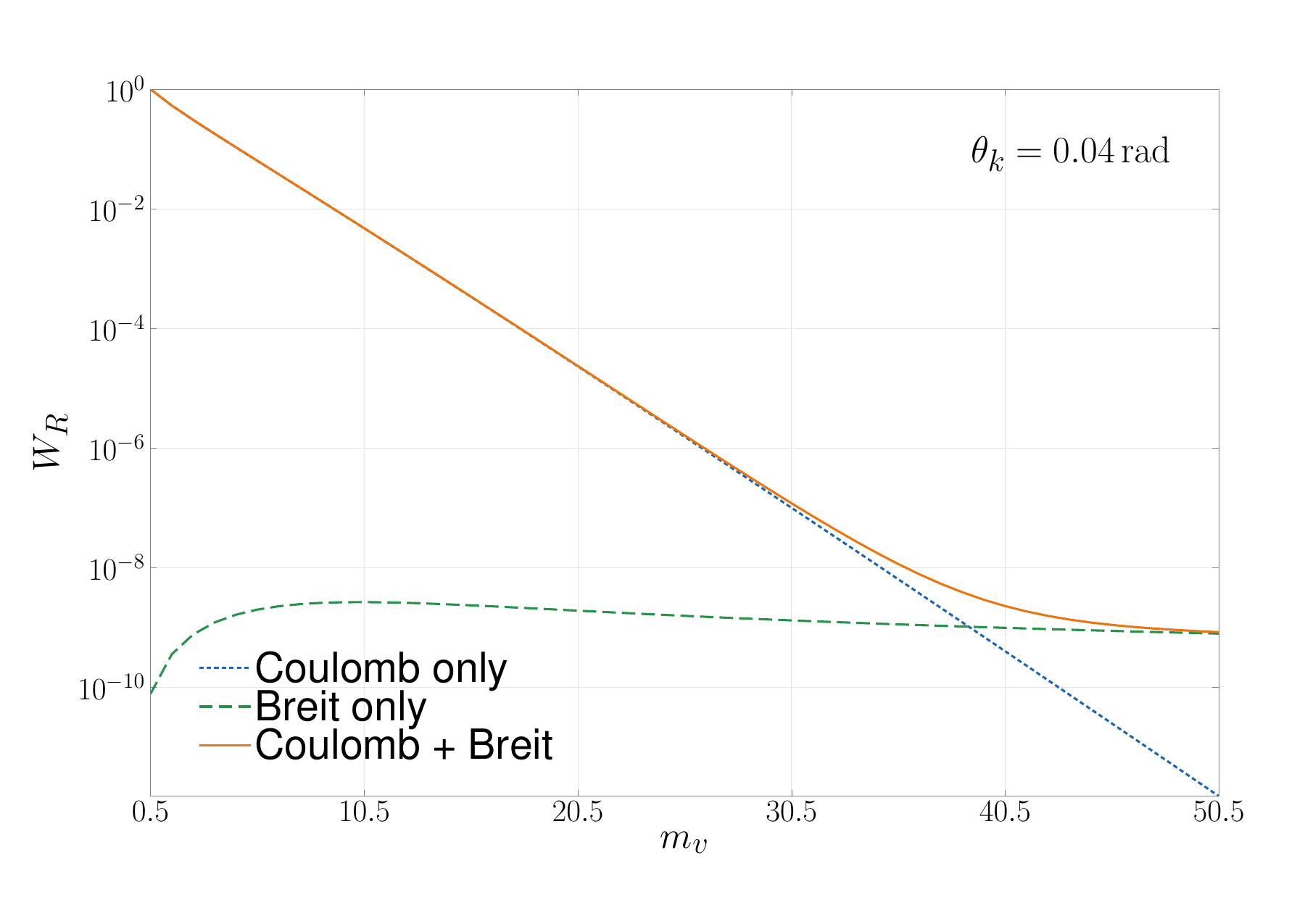}

    \caption{\raggedright Relative effective cross section $W_R(m_v)$ and its dependence on $m_v$ for the single-atom target case.
The kinetic energy of the initial vortex electron is $E_{\mathrm{kin}}=300~\mathrm{keV}$.
The conical angles of the initial vortex electron are $\theta_k=0.01$ rad, $0.02$ rad, $0.03$ rad, and $0.04$ rad respectively.
The helicities are $\lambda_k=\lambda_p=\frac{1}{2}$.}
    \label{fig:mvdep_thetak}
\end{figure}

We examine how the effective cross section $W_R$ depends on the TAM $m_v$ of the initial vortex electron.
This dependence can be influenced by the kinetic energy and the conical angle of the initial vortex electron.
We therefore present numerical results for $W_R(m_v)$ with varying kinetic energy $E_{\rm kin}$ and conical angle $\theta_k$.
Three cases are compared: the ``Coulomb only'' case, the ``Breit only'' case, and the ``Coulomb+Breit'' case.
The results are shown in FIG.~\ref{fig:mvdep_Ekin} and FIG.~\ref{fig:mvdep_thetak}.

\paragraph{Varying kinetic energy $E_{\rm kin}$}
Keeping the conical angle of the initial vortex electron constant at $\theta_k=0.01$ rad, we compare $W_R(m_v)$ for four different kinetic energies of the initial vortex electron: $E_{\rm kin}=100$ keV, $300$ keV, $500$ keV, and $700$ keV.
The helicities of both the incident and scattered electrons are set to $\frac{1}{2}$.
The results are shown in FIG.~\ref{fig:mvdep_Ekin}.
From the figure, we see that, for the pure Coulomb contribution, the effective cross section decreases rapidly with increasing $m_v$.
On a logarithmic scale, this decrease is nearly linear in $m_v$, indicating an approximately exponential suppression of the form $\propto 10^{-\alpha m_v}$ with $\alpha > 0$.
For the pure Breit contribution, the effective cross section also decreases with increasing $m_v$, except for a small region with $m_v\le 3.5$.
However, the decrease for the pure Breit contribution is far slower than that for the pure Coulomb contribution.
Since the ``Coulomb only'' case is much larger than the ``Breit only'' case at small $m_v$ values, the two lines intersect at a certain point.
This point acts as a threshold around which the Breit and Coulomb contributions are comparable in magnitude, and it divides $W_R(m_v)$ into two regimes.
For $m_v$ below this threshold, the ``Coulomb only'' and ``Coulomb+Breit'' curves coincide with each other, which means that the Coulomb contribution dominates the process, whereas for $m_v$ above it, the ``Breit only'' and ``Coulomb+Breit'' curves coincide with each other, which means that the Breit contribution dominates the process.
From the four sub-figures in FIG.~\ref{fig:mvdep_Ekin} with different initial $E_{\rm kin}$, we see that the threshold values are different.
The larger $E_{\rm kin}$ is, the larger the threshold is.
The thresholds in the four sub-figures are listed in TABLE~\ref{table_Ekin}.
\begin{table}[htbp]
  \centering
  \begin{tabular}{|c|c|c|c|c|}
    \hline
    Kinetic energy $E_{\rm kin}$ (keV) & 100 & 300 & 500 & 700 \\
    \hline
    The threshold of $m_v$ & 9.5 & 13.5 & 17.5 & 20.5 \\
    \hline
  \end{tabular}
  \caption{$m_v$ thresholds for different $E_{\rm kin}$}
  \label{table_Ekin}
\end{table}

\paragraph{Varying conical angle $\theta_k$}
Keeping the kinetic energy of the initial vortex electron constant at $E_{\rm kin}=300$ keV, we compare $W_R(m_v)$ for four different conical angles of the initial vortex electron: $\theta_p=0.01$ rad, $0.02$ rad, $0.03$ rad, and $0.04$ rad.
The helicities of both the incident and scattered electrons are set to $\frac{1}{2}$.
The results are shown in FIG.~\ref{fig:mvdep_thetak}.
From the sub-figures, we see that the behavior of the thresholds in this case, with varying conical angle $\theta_k$, is similar to that in the previous case with varying kinetic energy $E_{\rm kin}$.
As $\theta_k$ increases, the value of the special threshold, which divides $W_R(m_v)$ into two different regimes, also increases.
The thresholds in the four sub-figures are listed in TABLE~\ref{table_thetak}.
\begin{table}[htbp]
  \centering
  \begin{tabular}{|c|c|c|c|c|}
    \hline
    Conical angle $\theta_k$ (rad) & 0.01 & 0.02 & 0.03 & 0.04 \\
    \hline
    The threshold of $m_v$ & 13.5 & 22.5 & 30.5 & 39.5 \\
    \hline
  \end{tabular}
  \caption{$m_v$ thresholds for different $\theta_k$}
  \label{table_thetak}
\end{table}
There is still a small region at small $m_v$ values in which the effective cross section for ``Breit only'' case increases with increasing $m_v$.
This region also grows with larger $\theta_k$.
However, it does not affect the threshold effect, since the threshold is much larger than the $m_v$ values in this small region.

In summary, there exists a threshold value of $m_v$.
For $m_v$ exceeding this threshold, the Breit contribution dominates the scattering process.
This behavior departs from the conventional expectation that the Breit contribution can be safely neglected in elastic electron scattering from low-$Z$ atoms.
The direct reason for this behavior is that the decrease of the Breit contribution is far slower than that of the Coulomb contribution, which results in the fact that the Breit contribution is far larger than the Coulomb contribution at $m_v$ values larger than the threshold.
There is also an underlying reason.
In atomic physics, the Breit interaction encodes the magnetic and retardation effects in the electron--electron interaction, both of which are relativistic corrections to the Coulomb interaction.
Since the novel phenomenon identified here occurs at large TAM of the vortex electron---corresponding to a large intrinsic magnetic moment---we conclude that it is the magnetic interaction between the vortex electron and the bound atomic electron that results in the slow decrease of the Breit interaction and governs the scattering process when the TAM of the initial vortex electron is sufficiently large.

\subsection{$m_v$ dependence of effective cross section in Mesoscopic target case}
\begin{figure}[h!]
    \centering
        \centering
        \includegraphics[width=0.45\linewidth]{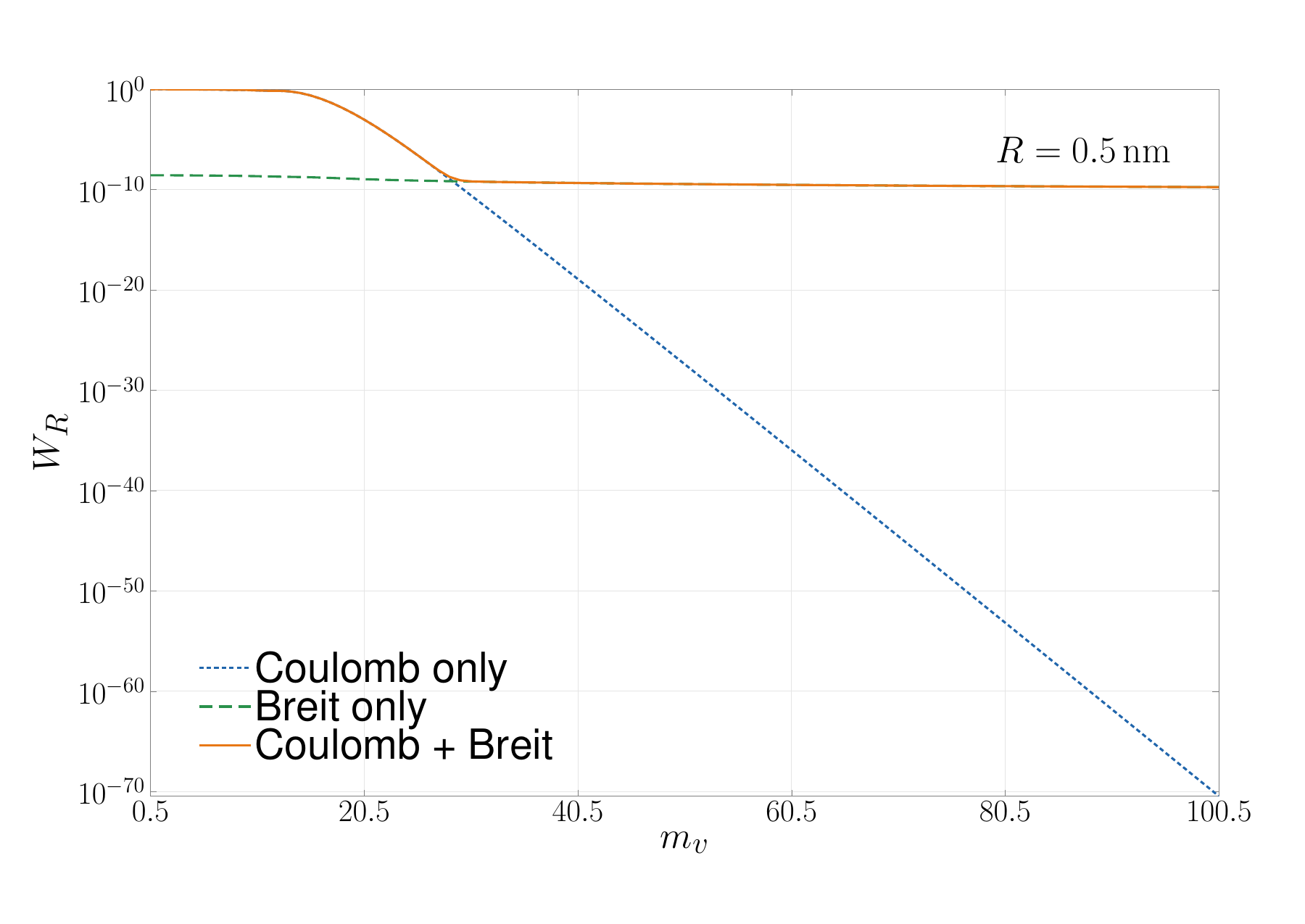}
        \includegraphics[width=0.45\linewidth]{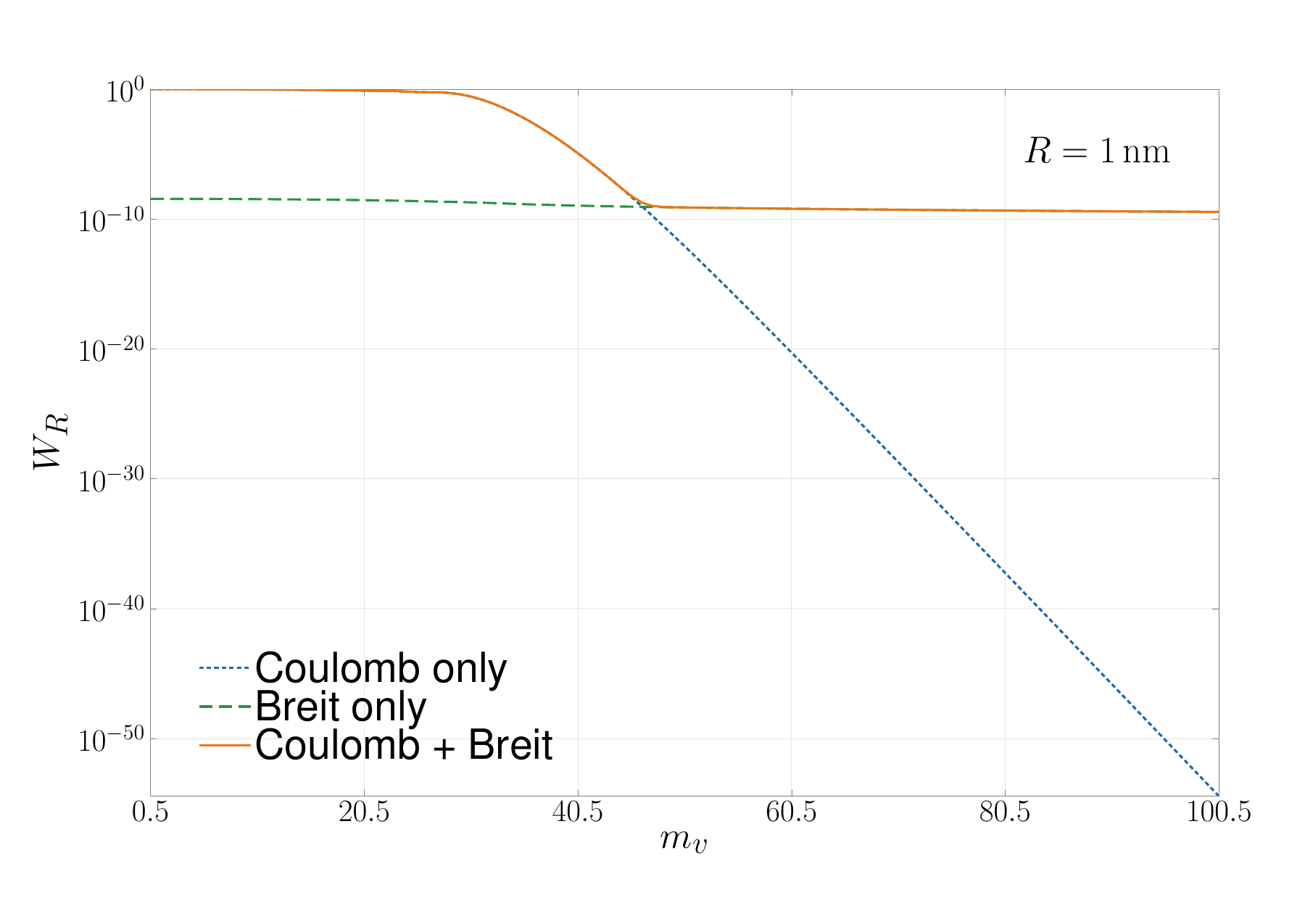}
        \includegraphics[width=0.45\linewidth]{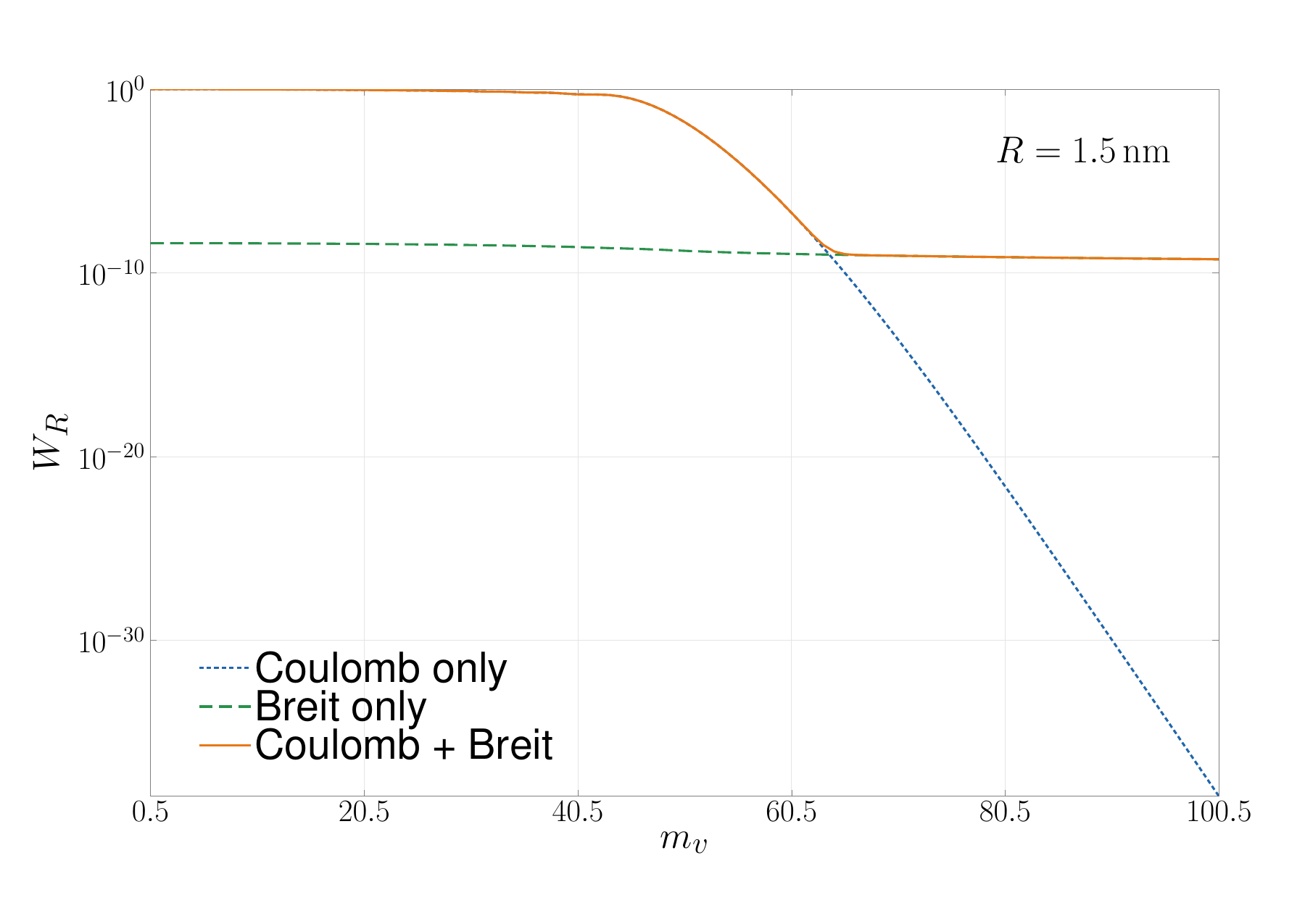}
        \includegraphics[width=0.45\linewidth]{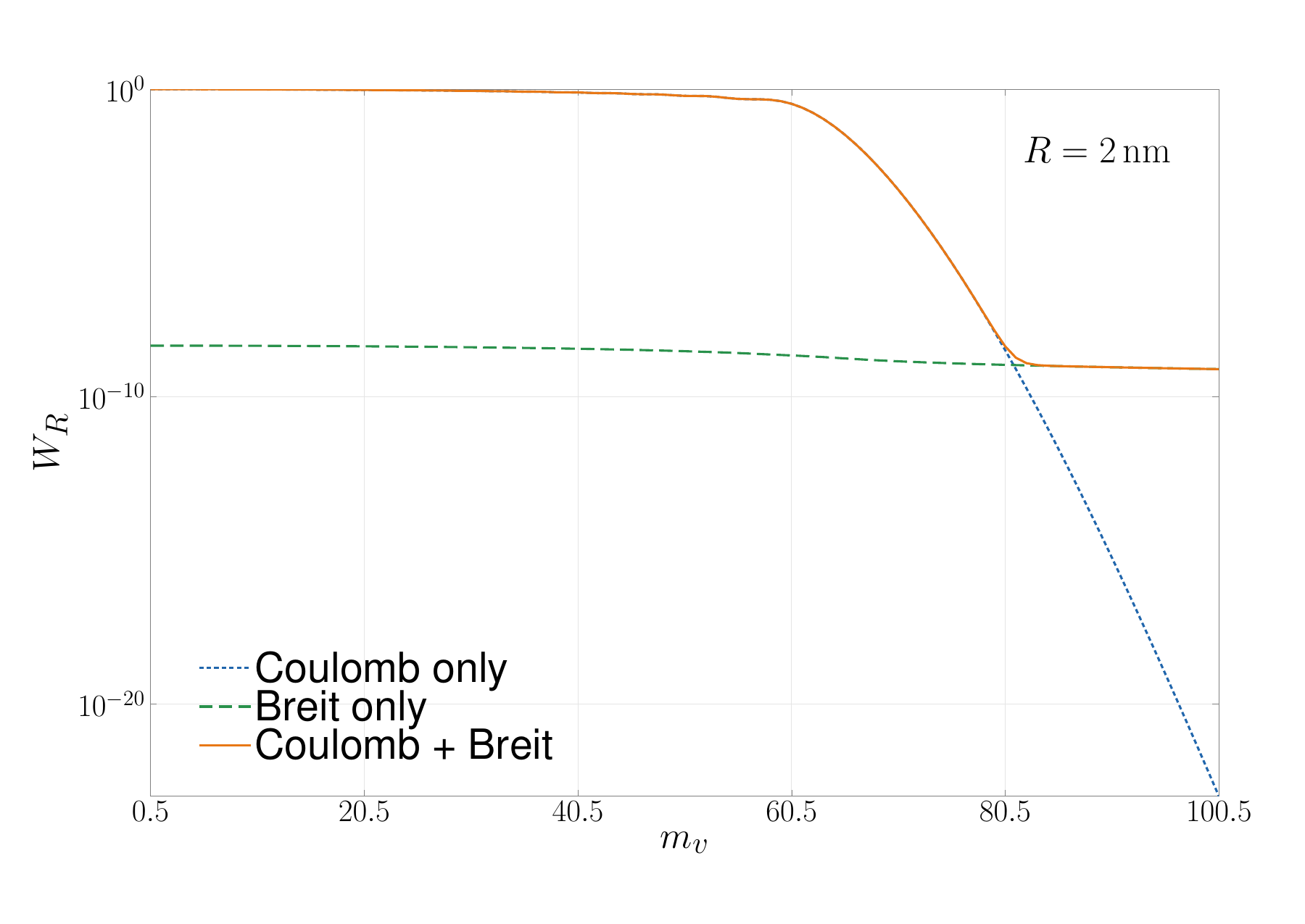}

    \caption{\raggedright Relative effective cross section $W_R(m_v)$ and its dependence on $m_v$ for the mesoscopic target case.
The kinetic energy of the initial vortex electron is $E_{\mathrm{kin}}=300~\mathrm{keV}$.
The conical angle of the initial vortex electron is $\theta_k=0.01$ rad.
The finite target sizes are $R_m=0.5$ nm, $1$ nm, $1.5$ nm, and $2$ nm.
The helicities are $\lambda_k=\lambda_p=\frac{1}{2}$.}
    \label{fig:mvdep_R}
\end{figure}
In this subsection, we present results for the mesoscopic target case and compare them with those for the single-atom target case.

To investigate whether the effect identified in the single-atom target case persists for the mesoscopic target case, which is more experimentally realistic, we focus on the $m_v$ dependence of the effective cross section.
The parameters for the initial vortex electron are $E_{\rm kin}=300$~keV and $\theta_k=0.01$~rad, which can be achieved experimentally in the transmission electron microscope nowadays.
The helicities of both the incident and scattered electrons are set to $\frac{1}{2}$.
To see how the finite target size influences the results, we examine four different target sizes: $R_m=0.5$ nm, $1$ nm, $1.5$ nm, and $2$ nm.
The $m_v$ dependence of the relative effective cross section $W_R(m_v)$ at different target sizes is shown in FIG.~\ref{fig:mvdep_R}.

The same qualitative behavior as in the single-atom target case is observed: there exists a threshold value of $m_v$ around which the Breit and Coulomb contributions are comparable in magnitude.
For $m_v$ below this threshold, the Coulomb contribution dominates, while for $m_v$ above it, the Breit contribution dominates.
Compared with the single-atom target case, the thresholds for the mesoscopic targets are shifted to larger $m_v$ values, given the same initial vortex electron state and observation angle (The second sub-figure in FIG.~\ref{fig:mvdep_Ekin} and the first sub-figure in FIG.~\ref{fig:mvdep_thetak}).
Moreover, from the sub-figures in FIG.~\ref{fig:mvdep_R}, we see that the threshold is influenced by the size of the target.
The larger the target radius, the larger the threshold.
The thresholds in the four sub-figures are listed in TABLE~\ref{table_R}.
\begin{table}[htbp]
  \centering
  \begin{tabular}{|c|c|c|c|c|}
    \hline
    Finite target size $R$ (nm) & 0.5 & 1 & 1.5 & 2 \\
    \hline
    Threshold of $m_v$ & 29.5 & 47.5 & 64.5 & 81.5 \\
    \hline
  \end{tabular}
  \caption{$m_v$ thresholds for different target sizes $R$}
  \label{table_R}
\end{table}
This is consistent with the fact that the effective differential cross section is independent of $m_v$ for a macroscopic target, where the TAM effect is averaged out.
As the target radius increases, the threshold shifts to ever larger $m_v$, and in the limit of an infinitely large target, the threshold diverges to infinity and the effect vanishes entirely.

\section{Conclusions}\label{section four}
We have studied elastic scattering of a vortex electron by a hydrogen
atomic target.
The Hamiltonian employed includes both the Coulomb and Breit
interactions, with the aim of demonstrating that the Breit interaction
cannot be neglected in electron scattering from low-$Z$ atomic targets
when the incident electron is in a vortex state.
We have presented the effective differential cross section as a function of the scattering angle of the outgoing electron.
Most importantly, we have shown how the effective cross section depends on the TAM $m_v$ of the initial vortex electron.
Two scenarios have been compared: one in which only the Coulomb
interaction contributes, and the other in which only the Breit interactions contribute.
Both the single-atom and mesoscopic target cases have been studied,
and the macroscopic target case has also been discussed.

Our results demonstrate that there exists a threshold value of $m_v$
around which the Breit and Coulomb contributions are comparable in
magnitude.
This threshold divides the scattering into two regimes.
Below the threshold, the Coulomb contribution dominates the process.
Above the threshold, the Breit contribution dominates, despite the
target being hydrogen.
This departs from the conventional expectation that the Breit
interaction can be safely neglected in elastic electron scattering
from low-$Z$ atoms.
The threshold is sensitive to the initial vortex electron parameters and the atomic target size; specifically, higher electron energy, larger conical angles, and larger target sizes all result in higher threshold.
Since the effective differential cross section decreases with increasing TAM,
the target size should be made as small as possible in experiments to
facilitate observation of this effect.
The above results show that, in applications of vortex electrons involving large TAM, the Breit interaction must be taken into account in calculations. These results provide important theoretical support for experiments on vortex electron applications.

\section*{Acknowledgments}
%%%%%%%%%%%%%%%%%%%%%%%%%%
This work was supported by Grant No.\,NSFC-12447117.

\begin{appendices}
\section{Effective differential cross section}\label{appendix a}
In this paper, we use initial Bessel vortex electron states with fixed TAM and helicity.
Their TAM equals $\hbar m_v$, where $m_v$ takes half-integer values.
Since the conventional definition of a cross section is not directly applicable for this kind of initial state, we construct an effective differential cross section.
When $m_v=\pm\frac{1}{2}$ and $\kappa$ tends to zero, the state tends to a plane-wave state, apart from a constant factor.
Thus, in this limit, the effective differential cross section should tend to the differential cross section for the initial plane-wave case.
We will construct an effective differential cross section that is consistent with this limiting behavior.

For $\lambda_k = \frac{1}{2}$, we obtain
\begin{align}\label{bessel +}
    \psi^B_+(\vb r) &= \int \frac{d^2 \vec k_\perp}{(2\pi)^2} \, \frac{1}{\sqrt{2E_k}} a_{\kappa m_v}(\vec k_\perp) \, \mathrm{e}^{i\vb k \cdot \vb r}\nonumber\\
    &\quad \times
    \begin{pmatrix}
        \sqrt{E_k + m_e} \cos\frac{\theta_k}{2} \, \mathrm{e}^{-i\frac{\phi_k}{2}} \\
        \sqrt{E_k + m_e} \sin\frac{\theta_k}{2} \, \mathrm{e}^{i\frac{\phi_k}{2}} \\
        \sqrt{E_k - m_e} \cos\frac{\theta_k}{2} \, \mathrm{e}^{-i\frac{\phi_k}{2}} \\
        \sqrt{E_k - m_e} \sin\frac{\theta_k}{2} \, \mathrm{e}^{i\frac{\phi_k}{2}}
    \end{pmatrix} \nonumber\\
    &=  \frac{i^{-\frac{1}{2}}}{\sqrt{2E_k}}\sqrt{\frac{\kappa}{2\pi}} \mathrm{e}^{ik_z z}\nonumber\\
    &\quad \times
    \begin{pmatrix}
        \sqrt{E_k + m_e} \cos\frac{\theta_k}{2} \, J_{m_v-\frac{1}{2}}(\kappa r_\perp) \, \mathrm{e}^{i(m_v-\frac{1}{2})\phi_r} \\
        i \sqrt{E_k + m_e} \sin\frac{\theta_k}{2} \, J_{m_v+\frac{1}{2}}(\kappa r_\perp) \, \mathrm{e}^{i(m_v+\frac{1}{2})\phi_r} \\
        \sqrt{E_k - m_e} \cos\frac{\theta_k}{2} \, J_{m_v-\frac{1}{2}}(\kappa r_\perp) \, \mathrm{e}^{i(m_v-\frac{1}{2})\phi_r} \\
        i \sqrt{E_k - m_e} \sin\frac{\theta_k}{2} \, J_{m_v+\frac{1}{2}}(\kappa r_\perp) \, \mathrm{e}^{i(m_v+\frac{1}{2})\phi_r}
    \end{pmatrix} .
\end{align}
When $m_v=\frac{1}{2}$ and $\kappa$ tends to zero,
\begin{align}\label{t+}
    \psi ^B_+\rightarrow i^{-\frac{1}{2}}\sqrt{\frac{\kappa}{2\pi}}\psi ^P_+,
\end{align}
with the momentum $\vb k$ pointing along the $z$ direction.
For $\lambda_k = -\frac{1}{2}$, we obtain
\begin{align}\label{bessel -}
    \psi^B_-(\vb r) &= \int \frac{d^2 \vec k_\perp}{(2\pi)^2} \, \frac{1}{\sqrt{2E_k}}a_{\kappa m_v}(\vec k_\perp) \, \mathrm{e}^{i\vb k \cdot \vb r}\nonumber\\
    &\quad \times
    \begin{pmatrix}
        -\sqrt{E_k + m_e} \sin\frac{\theta_k}{2} \, \mathrm{e}^{-i\frac{\phi_k}{2}} \\
        \sqrt{E_k + m_e} \cos\frac{\theta_k}{2} \, \mathrm{e}^{i\frac{\phi_k}{2}} \\
        \sqrt{E_k - m_e} \sin\frac{\theta_k}{2} \, \mathrm{e}^{-i\frac{\phi_k}{2}} \\
        -\sqrt{E_k - m_e} \cos\frac{\theta_k}{2} \, \mathrm{e}^{i\frac{\phi_k}{2}}
    \end{pmatrix} \nonumber\\
    &=  \frac{i^{-\frac{1}{2}}}{\sqrt{2E_k}}\sqrt{\frac{\kappa}{2\pi}}\mathrm{e}^{ik_z z}\nonumber\\
    &\quad \times
    \begin{pmatrix}
        -\sqrt{E_k + m_e} \sin\frac{\theta_k}{2} \, J_{m_v-\frac{1}{2}}(\kappa r_\perp) \, \mathrm{e}^{i(m_v-\frac{1}{2})\phi_r} \\
        i \sqrt{E_k + m_e} \cos\frac{\theta_k}{2} \, J_{m_v+\frac{1}{2}}(\kappa r_\perp) \, \mathrm{e}^{i(m_v+\frac{1}{2})\phi_r} \\
        \sqrt{E_k - m_e} \sin\frac{\theta_k}{2} \, J_{m_v-\frac{1}{2}}(\kappa r_\perp) \, \mathrm{e}^{i(m_v-\frac{1}{2})\phi_r} \\
        -i \sqrt{E_k - m_e} \cos\frac{\theta_k}{2} \, J_{m_v+\frac{1}{2}}(\kappa r_\perp) \, \mathrm{e}^{i(m_v+\frac{1}{2})\phi_r}
    \end{pmatrix} .
\end{align}
When $m_v=-\frac{1}{2}$ and $\kappa$ tends to zero,
\begin{align}\label{t_}
    \psi ^B_-\rightarrow i^{\frac{1}{2}}\sqrt{\frac{\kappa}{2\pi}}\psi ^P_-,
\end{align}
with the momentum $\vb k$ pointing along the $z$ direction.
The normalization condition for both kinds of states is
\begin{align}
    \braket{\psi ^B}{\psi ^B}=\frac{R_tL}{\pi },
\end{align}
where $R_t$ is the infinite transverse length, $L$ is the infinite longitudinal length, and the integral
\[
\int_0^{R_t}d\rho \,\kappa \rho J^2_m(\kappa \rho)\approx \frac{R_t}{\pi}
\]
has been used.

We can define an effective differential cross section:
\begin{align}\label{ecs}
    d\hat {\sigma}^B=\frac{d\Gamma}{\hat j_z},
\end{align}
where $d\Gamma$ is the differential transition rate and $\hat j_z$ is the effective incident current density.
According to Fermi's golden rule, the differential transition rate can be calculated as
\begin{align}
    d\Gamma=2\pi\delta(E_p-E_k)\frac{|{\cal M}^B|^2}{\braket{\psi^B_i}{\psi^B_i} \braket{\psi^P_f}{\psi^P_f}}\frac{Vd^3\vb p}{(2\pi)^3},
\end{align}
where $V$ is the volume of the three-dimensional space,
\begin{align}
    \braket{\psi ^B_i}{\psi ^B_i}=\frac{ R_tL}{\pi },
\end{align}
and
\begin{align}
    \braket{\psi ^P_f}{\psi ^P_f}=V=\pi R_t^2L.
\end{align}
After performing the integration that eliminates the delta function, we get
\begin{align}
    \frac{d \Gamma}{d\Omega} = \frac{\pi }{ R_tL}\frac{pE_k}{4\pi^2}\,|{\cal M}^B|^2=\frac{pE_k}{4\pi R_tL}\, |{\cal M}^B|^2.
\end{align}
The effective incident current density can be chosen as the average current density in the transverse plane:
\begin{align}
    \hat j_z=N_j\frac{1\cdot v_B}{L}\frac{1}{\pi R_t^2}=\frac{N_jv_B}{\pi R_t^2L},
\end{align}
where $N_j$ is a constant factor to be adjusted so that Eq.~\eqref{ecs} reproduces the differential cross section for the initial plane-wave case in the limit $m_v=\pm \frac{1}{2}$ and $\kappa \rightarrow 0$.
With this effective incident current density, we obtain
\begin{align}
    \frac{d\hat {\sigma}^B}{d\Omega}=\frac{ R_t p E_k}{4 N_j v_B}|{\cal M}^B|^2.
\end{align}
In the limit $m_v=\frac{1}{2}$ and $\kappa$ tends to zero, this should reproduce the result for the differential cross section in the initial plane-wave case:
\begin{align}\label{p}
    \frac{d\sigma ^P}{d\Omega}=\frac{pE_k}{4\pi^2 v_k}|{\cal M}^P|^2.
\end{align}
According to Eq.~\eqref{t+} and Eq.~\eqref{t_}, we have
\begin{align}
    |{\cal M}^B|^2\rightarrow \frac{\kappa}{2\pi}|{\cal M}^P|^2
\end{align}
in the limit $m_v=\frac{1}{2}$ and $\kappa$ tends to zero.
Thus
\begin{align}
    \frac{d\hat {\sigma}^B}{d\Omega}\rightarrow \frac{\kappa R_t p E_k}{8\pi N_j v_k}|{\cal M}^P|^2.
\end{align}
Comparing with Eq.~\eqref{p}, this implies that $N_j=\frac{\pi \kappa R_t}{2}$.
Thus a correct effective differential cross section for the initial vortex-state case should be
\begin{align}
    \frac{d\hat {\sigma}^B}{d\Omega}=\frac{2}{\pi \kappa R_t}\frac{ R_t p E_k}{4 v_B}|{\cal M}^B|^2=\frac{  E_k^2}{2\pi  \kappa \cos \theta_k}|{\cal M}^B|^2.
\end{align}

\end{appendices}

%\bibliography{paper-version/refs.bib}{}
% \bibliographystyle{JHEP}
%\bibliographystyle{utphys28mod}
% \bibliographystyle{apalike}
\bibliographystyle{utphys} % 选择参考文献样式
\bibliography{refs} % 指定.bib文件名（不带扩展名）

\end{document}